# Boson peak and medium-range elastic heterogeneity in calcium silicate hydrate probed by terahertz spectroscopy and low-temperature calorimetry

Xiangyu Li[a,*], Ying Chen[a], Jipeng Luo[b], Ya Chen[c], Linhao Wang[d], Lidan Tian[d], Gan Ding[e], Zeyu Lu[f], Zhangli Hu[f], Biqin Dong[g], Yue Li[a], Zongjin Li[h]

[a] College of Architecture and Civil Engineering, Beijing University of Technology, Beijing 100124, China

[b] Dalian Institute of Chemical Physics, Chinese Academy of Sciences, Dalian 116023, China

[c] School of Environment and Safety Engineering, North University of China, Taiyuan 030051, China

[d] Department of Civil Engineering, Taiyuan University of Technology, Taiyuan 030024, China

[e] State Key Laboratory of Nonlinear Mechanics, Institute of Mechanics, Chinese Academy of Sciences, Beijing, China

[f] School of Materials Science and Engineering, Southeast University, Nanjing 211189, China

[g] College of Civil and Transportation Engineering, Shenzhen University, Shenzhen 518060, China

[h] Faculty of Innovation Engineering, Macau University of Science and Technology, Macau SAR, China

## Abstract

The boson peak (BP), a universal low-frequency vibrational anomaly of disordered solids, has been predicted but not systematically characterized in calcium silicate hydrate (C-S-H), the binding phase of hardened cement. Building on a preliminary terahertz survey of the same samples, we characterize the BP across five Ca/Si ratios (0.5–1.7) using terahertz time-domain spectroscopy (THz-TDS) and low-temperature calorimetry, two probes of vibrational dynamics that complement the static, time-averaged picture given by conventional crystallographic and spectroscopic methods. After Bruggeman correction for crystalline impurities, both probes locate the BP near 1 THz; they agree on frequency but diverge in intensity. The terahertz integrated spectral weight and the calorimetric $C_p/T^3$ peak both fall monotonically with Ca/Si, whereas the apparent terahertz peak height is maximal at Ca/Si ≈ 1.0, where the damping is low and the oscillator strength still substantial. This decoupling marks a structural crossover between silicate-chain depolymerization and interlayer calcium filling. From the BP we obtain a medium-range dynamical correlation length of order 1 nm (0.3–2 nm; not directly accessible to conventional cement-science techniques) and a coherent-potential elastic-heterogeneity parameter that decreases from $\gamma$ = 0.98 to 0.48 as Ca/Si rises; the Debye-normalized BP frequency ($\nu_{BP}/\nu_D$ = 0.15–0.17) places C-S-H within the range reported for silicate glasses. Because $\gamma$ governs the distribution of energy barriers for local structural rearrangements, it provides a quantitative, composition-resolved descriptor relevant to the intrinsic creep and thermal transport of C-S-H, linking nanoscale vibrational dynamics to the macroscopic durability of concrete. More broadly, the dual-probe boson-peak approach is transferable to other amorphous solids, including the amorphous supplementary cementitious materials (slag, fly ash, calcined clays) of low-carbon cements.

## 1. Introduction

Calcium silicate hydrate (C-S-H) is the primary binding phase formed during Portland cement hydration, typically accounting for 60–70% of the hydration products and governing both strength and durability of concrete [1,2]. Its structure is often described as a defective tobermorite-like layered silicate, with the calcium-to-silicon molar ratio (Ca/Si) ranging from about 0.6 to 2.0 depending on the cementitious system and the degree of hydration [3,4]. Despite decades of effort, establishing quantitative structure-property relationships for C-S-H remains difficult, largely because the material is X-ray amorphous yet nanocrystalline and compositionally variable at the nanometer scale. Crucially, the long-term performance of concrete is set less by the average structure of C-S-H than by the nanoscale spatial heterogeneity of its stiffness: this elastic heterogeneity governs the distribution of energy barriers for the stress-driven local rearrangements that produce creep, the time-dependent deformation that dominates the long-term deflection and prestress loss of concrete structures [5], and it limits the low-frequency vibrations that carry heat through the binder. Thermal transport itself spans an unusually wide range of service conditions, from cryogenic infrastructure, where the inner walls of liquefied-natural-gas and liquid-hydrogen storage tanks operate near 110 K and below, to massive and extreme-climate structures in which steep thermal gradients drive thermal stress and cracking. Yet this controlling elastic heterogeneity has remained beyond direct experimental reach.

C-S-H has been characterized in detail at the atomic and mesoscale levels. At the atomic, short-range-order (SRO) level, solid-state NMR and vibrational spectroscopy have mapped the local coordination of silicon and calcium, the connectivity of silicate tetrahedra, and the evolution of chain length with Ca/Si [6,7]. At the mesoscale, small-angle scattering and electron microscopy have characterized the layered morphology, globular packing, and pore structure [8,9]. The decisive regime, however, lies in between, at the medium range of roughly 0.5 to 5 nm, where the partial ordering of the silicate framework within a glass-like matrix [10,11] is expressed mechanically as a spatial fluctuation of stiffness, namely the elastic heterogeneity. This regime is the hardest to reach: Fourier-transform infrared (FTIR) and Raman spectroscopy probe primarily SRO bond vibrations, X-ray diffraction (XRD) yields broad, poorly resolved patterns for a turbostratically disordered phase, and NMR provides only spatially averaged coordination environments, all reporting a static, time-averaged structure; nanoindentation, in turn, returns only volume- and orientation-averaged moduli. Capturing the elastic heterogeneity instead calls for a probe of the low-frequency vibrational dynamics in which it is encoded.

In the physics of disordered solids, the boson peak (BP) is the spectroscopic fingerprint of exactly this elastic heterogeneity, and one of the most universal and intensively studied phenomena of the glassy state, although it is essentially unknown in cement science. The BP manifests as an excess of vibrational states relative to the Debye model prediction at low frequencies, appearing as a broad hump in the reduced vibrational density of states $g(\omega)/\omega^2$ at frequencies typically between 0.3 and 3 THz [12,13]. Its physical origin, while still debated, is generally associated with quasi-localized vibrations arising from nanoscale elastic heterogeneity; the characteristic frequency scales as $\omega_{BP} \sim v/\xi$, where v is the transverse sound velocity and $\xi$ a structural correlation length on the order of 1 nm [14–16]. In silicate glasses, the structural analogues of C-S-H, the BP has been extensively studied and shown to encode information about network connectivity, modifier content, and thermal history [17,18]. Pesina et al. [19] established a direct quantitative link between BP frequency and medium-range structural parameters in zinc sodium silicate glasses. For C-S-H, Abdolhosseini Qomi et al. [20] computed the vibrational density of states by molecular dynamics for 150 models spanning Ca/Si = 1.1–2.1 and located the BP within a broad ~1.2–2.5 THz range that shifts with composition, arising from collective Ca-O and $SiO_4$ vibrations, rather than at a single fixed frequency. That prediction, however, has not been tested experimentally.

Two complementary techniques are well suited for detecting the BP in C-S-H. Terahertz time-domain spectroscopy (THz-TDS) measures the complex dielectric response in the 0.1–10 THz range, a window that coincides with the predicted BP frequency. Unlike conventional IR or Raman spectroscopy, THz-TDS is sensitive to low-frequency collective excitations and dipolar fluctuations in disordered systems [21–23], and coherent detection provides both amplitude and phase information without Kramers-Kronig transformation. Low-temperature heat capacity measurement offers an independent, thermodynamic route: in amorphous solids, the ratio $C_p/T^3$ displays a broad peak at a temperature corresponding to the BP energy scale, reflecting the integrated excess density of states over all vibrational modes regardless of their dipole activity [24,25]. The two probes couple to the BP through different mechanisms, THz-TDS weights dipole-active modes, while calorimetry integrates over all modes, and their combination can both confirm the BP and reveal aspects of the vibrational spectrum that neither technique alone can access. Although applied here to C-S-H, this dual-probe strategy is general to disordered solids, and within cement science it extends naturally to the amorphous supplementary cementitious materials (fly ash, slag, calcined clays) increasingly used in low-carbon binders, including limestone calcined clay ($LC^3$) cements.

Building on a preliminary terahertz survey of this sample series that first noted boson-peak-like features [26], here we report a systematic experimental characterization of the BP in synthetic C-S-H across five Ca/Si ratios (0.5, 0.8, 1.0, 1.4, and 1.7), using THz-TDS and low-temperature calorimetry as independent probes. The intrinsic dielectric response of C-S-H is isolated from crystalline impurities ($Ca(OH)_2$ and $CaCO_3$) through effective medium theory, and the BP is parameterized by fitting a damped harmonic oscillator (DHO) model to the normalized dielectric loss spectra. The calorimetric $C_p/T^3$ data confirm the presence of excess low-frequency vibrational states at a consistent energy scale. We find that both the calorimetric BP intensity and the THz integrated spectral weight decrease monotonically with increasing Ca/Si, yet the apparent THz peak height reaches a maximum at Ca/Si ≈ 1.0, where the vibrational damping is low while the oscillator strength remains substantial. This decoupling is traced to the structural crossover from silicate chain depolymerization to interlayer calcium filling, and reveals that Ca/Si ≈ 1.0 represents a state of maximum vibrational coherence for the low-frequency excess modes. From the boson peak we obtain a coherent-potential parameter $\gamma$ that quantifies the medium-range elastic heterogeneity of C-S-H directly, to our knowledge the first experimental measure of this quantity, together with a dynamical correlation length of order 1 nm (0.3–2 nm); both pertain to a scale not directly accessible to the characterization techniques conventional in cement science.

## 2. Materials and methods

### 2.1. Synthesis of C-S-H

C-S-H samples with nominal Ca/Si molar ratios of 0.5, 0.8, 1.0, 1.4, and 1.7 were synthesized by the direct reaction of CaO and $SiO_2$ in water (the lime-silica, or pozzolanic, route) [27,28]. Reactive CaO, obtained by calcining reagent-grade $CaCO_3$ at 1000 °C for 2 h, was stored in a desiccator and handled under $N_2$ to limit rehydration and carbonation before being mixed with fumed $SiO_2$ (specific surface area ~200 $m^2$/g) at the prescribed molar ratio. The mixed powders were dispersed in $CO_2$-free (decarbonated) deionized water (water/solid ≈ 50 by mass) and stirred continuously at room temperature for 28 days under $N_2$ atmosphere to prevent carbonation. The resulting suspensions were centrifuged and washed three times with deionized water. Free water was removed by solvent exchange with isopropanol, followed by vacuum drying at 60 °C for 48 h. The dried products were ground gently in an agate mortar and stored in sealed vials under $N_2$ prior to characterization. Because phase-pure C-S-H is obtained by this route only for Ca/Si$\lesssim$1.5 [27,28], minor $Ca(OH)_2$ and trace $CaCO_3$ are expected at the highest nominal ratios; the actual Ca/Si of the C-S-H phase and the crystalline impurity contents were quantified by X-ray fluorescence (XRF), Thermogravimetric analysis (TGA), and XRD (Section 2.2) and accounted for in all subsequent analysis (Section 2.5).

### 2.2. Characterization of phase composition

XRD patterns were recorded on a Rigaku SmartLab diffractometer (Cu Kα radiation, 40 kV, 30 mA) over 2θ = 5°–90° at 5°/min. FTIR spectra were collected on a Thermo Fisher Nicolet iS20 spectrometer using KBr pellets (4000–400 $cm^{-1}$, 4 $cm^{-1}$ resolution). TGA was performed on a Hitachi STA200 analyzer from 30 to 800 °C at 10 °C/min under $N_2$ flow (10 mL/min). The mass fractions of $Ca(OH)_2$ and $CaCO_3$ were calculated from the mass losses in the 400–500 °C and 600–800 °C intervals, respectively [29]. Oxide compositions were determined by XRF (PANalytical Axios), and the actual Ca/Si ratios of the C-S-H phase were computed after subtracting the calcium contributed by $Ca(OH)_2$ and $CaCO_3$. True densities were measured on the dried powders by helium pycnometry (Micromeritics AccuPyc II 1340). Particle size distributions were obtained by laser diffraction (TopSizer Plus) in wet dispersion mode.

**2.3. THz-TDS measurements**

C-S-H powders were diluted with cyclic olefin copolymer (COC) and pressed into pellets for transmission measurements, following the THz-TDS approach we applied to tricalcium silicate [30]. COC was chosen over the more commonly used polyethylene (PE) because of its lower absorption coefficient and flatter spectral response across the 0.1–4 THz range, which extends the usable bandwidth for weakly absorbing analytes. It also produces mechanically stable pellets with flat, parallel surfaces. Three parameters govern the quality of THz transmission pellets: the C-S-H-to-COC mass ratio, the total pellet mass, and the pressing force. Higher C-S-H loading increases the signal from the analyte but also raises absorption losses, narrowing the usable spectral bandwidth; conversely, excessive dilution weakens the C-S-H spectral features relative to noise. To identify the combination that maximizes the effective spectral bandwidth while maintaining adequate signal-to-noise ratio, we performed a systematic optimization using an orthogonal experimental design ($L_9(3^4)$) with three levels for each factor. The effective bandwidth was defined as the frequency range over which the transmitted power remained at least 10 dB above the noise floor. The optimized parameters, C-S-H: COC mass ratio of 1:19 (5 wt% C-S-H), total mass 60 mg, pressing force 5 MPa held for 5 min in a 13 mm die, yielded pellets approximately 0.7 mm thick with an effective bandwidth extending from 0.1 to beyond 5.0 THz, adequate for the present study. Among the three factors, the C-S-H loading had the strongest influence on bandwidth: increasing the loading from 5% to 15% reduced the upper frequency limit from >3.0 THz to ~1.8 THz due to the high intrinsic absorption of C-S-H in the THz range. The two powders were hand-mixed in an agate mortar until visually homogeneous before pressing.

Transmission THz-TDS measurements were carried out on an Advantest TAS 7500SU spectrometer (spectral coverage 0.1–7.0 THz, dynamic range 57 dB, frequency resolution 7.6 GHz). Each pellet was centered in the collimated beam path, and the transmitted time-domain waveform was recorded. An air reference was measured under identical conditions.

At least five repeat scans were averaged for each sample to improve the signal-to-noise ratio. The complex refractive index was extracted from the transmission function $T(\omega) = E_{sam}(\omega)/E_{ref}(\omega)$ following the standard analytical procedure [31]:

$$n(\nu) = 1+(c\cdot\Delta\varphi(\nu))/(2\pi\nu d)$$

$$\kappa(\nu) = -c/(2\pi\nu d)\ln[(n+1)^2/4n\cdot|T(\nu)|]$$

where $\Delta\varphi$ is the phase difference between sample and reference, d the pellet thickness, c the speed of light, and $\nu$ the frequency. The absorption coefficient $\alpha = 4\pi\nu\kappa/c$, and the complex dielectric function $\varepsilon(\nu) = \varepsilon'(\nu) + i\varepsilon''(\nu)$ was obtained from $\varepsilon' = n^2-\kappa^2$ and $\varepsilon'' = 2n\kappa$.

**2.4. Low-temperature heat capacity**

Heat capacity was measured using a Quantum Design Physical Property Measurement System (PPMS) with the relaxation method over 2–300 K [32]. Because C-S-H is a fine powder, the sample was mixed with high-purity copper shavings to improve thermal conductivity, and the mixture was loaded into a high-purity copper cup (3 mm diameter, 0.025 mm wall thickness). The cup was sealed by pressing its side wall and then compressed with a pellet press to form a disc 1–3 mm in height. After mounting, the sample chamber was evacuated at room temperature to a high vacuum of $\sim 1.2 \times 10^{-3}$ Pa before cooling, and the specimen was then cooled from 300 K to 2 K in three staged ramps (10, 5 and 1 K $min^{-1}$ over the 300–20, 20–10 and 10–2 K intervals, respectively). Below 100 K, measurement points were spaced logarithmically (72 points); above 100 K, measurements were taken at 10 K intervals (21 points). The accuracy of the heat capacity measurement was verified by measuring standard reference materials, high-purity copper rod (>99.999%), $\alpha$-$Al_2O_3$ rod (SRM-720), and benzoic acid powder (SRM-39j), confirming standard uncertainties within ±3% below 20 K and ±1% over 20–300 K. For Ca/Si = 1.7, the measurement was performed twice on independently prepared specimens; the second measurement was used for all analyses (see Section 3.4 for discussion). To examine the BP, the data are plotted as $C_p/T^3$ versus T; in a Debye solid this ratio would approach a constant at low temperature, and any excess above that baseline signals the presence of additional low-frequency vibrational states.

**2.5. Correction for crystalline impurities**

Because the C-S-H samples contain minor amounts of $Ca(OH)_2$ and $CaCO_3$ (quantified by TGA, Section 2.2), the measured spectra of the composite pellets do not represent the intrinsic response of C-S-H alone. Both the THz dielectric data and the low-temperature heat capacity data were corrected using independently measured reference data for pure $Ca(OH)_2$ and $CaCO_3$.

*THz dielectric correction.* Reference dielectric functions of pure $Ca(OH)_2$ and $CaCO_3$ were measured on pellets prepared by the same COC dilution procedure. The intrinsic dielectric function of C-S-H was then extracted using the Bruggeman effective medium theory (EMT); effective-medium modelling of THz dielectric data has previously been applied to hardened cement paste [33]. Given the volume fractions $f_i$ determined from TGA and pycnometry, the Bruggeman mixing rule reads

$$\sum_i f_i (\varepsilon_i - \varepsilon_{eff})/(\varepsilon_i + 2\varepsilon_{eff}) = 0$$

where $\varepsilon_i$ denotes the dielectric function of the i-th component (C-S-H, $Ca(OH)_2$, $CaCO_3$, COC, and air voids) and $\varepsilon_{eff}$ the measured effective value. The dielectric function of C-S-H was solved numerically from this implicit equation at each frequency point. The Bruggeman model was selected over alternatives (Maxwell-Garnett, Wiener bounds) because it treats all phases symmetrically and is better suited for composites without a clear matrix-inclusion topology. The intrinsic $\varepsilon''$ spectra of crystalline $Ca(OH)_2$ and $CaCO_3$ in the BP frequency range (0.5–2.5 THz) are featureless; neither phase exhibits a peak in $\varepsilon''(\nu)/\nu$ (Fig. A.1). Given their low mass fractions ($Ca(OH)_2$ < 9.2%, $CaCO_3$ < 2%), the crystalline contribution to the composite $\varepsilon''$ at 1 THz is less than 3% of the C-S-H signal. Consequently, the EMT correction has minimal effect on the boson peak parameters, shifting $\nu_0$ by less than 2% and S by less than 5%. Perturbing the impurity volume fractions by ±20% in the EMT does not alter the non-monotonic trend in peak height, confirming that the BP signal is intrinsic to C-S-H.

*Heat capacity correction.* The heat capacity of pure $Ca(OH)_2$ and $CaCO_3$ powders was measured independently on the same PPMS instrument over the same temperature range (2–300 K). The intrinsic C-S-H heat capacity was obtained by subtracting the crystalline contributions according to their mass fractions:

$$C_{p,CSH} = (C_{p,measured}-\varphi_{CH}\cdot C_{p,CH}-\varphi_{CC}\cdot C_{p,CC})/(1-\varphi_{CH}-\varphi_{CC})$$

where $\varphi_{CH}$ and $\varphi_{CC}$ are the mass fractions of $Ca(OH)_2$ and $CaCO_3$ from TGA. Both crystalline phases are well-ordered solids that obey the Debye $T^3$ law at low temperatures, contributing a monotonic background to $C_p/T^3$ without producing any peak feature. The correction altered the $C_p/T^3$ peak heights by only 1.9–4.9% (smallest at Ca/Si = 0.8, largest at Ca/Si = 0.5 where the combined impurity fraction is highest), while peak positions and the compositional trend remained unchanged (Fig. A.2). A sensitivity analysis confirmed that even a ±30% uncertainty in the TGA-derived phase fractions changes the corrected peak heights by less than 4%, well within the measurement uncertainty. All heat capacity results reported below use the corrected data.

**2.6. Boson peak analysis**

***2.6.1. Frequency normalization of dielectric loss***

In disordered solids, the reduced vibrational density of states $g(\omega)/\omega^2$ exhibits the BP as a broad hump. For dielectric spectra, the analogous representation is the frequency-normalized dielectric loss $\varepsilon''(\nu)/\nu$, which removes the trivial linear frequency dependence expected from Debye relaxation and highlights any excess low-frequency absorption [34,35]. The same feature is equally visible as a shoulder in the absorption coefficient, whose normalized form $\alpha(\nu)/\nu^2$ carries the same boson-peak lineshape as $\varepsilon''(\nu)/\nu$; we adopt $\varepsilon''(\nu)/\nu$ as the more direct dielectric observable [34,35]. All BP analyses in this work are performed on the $\varepsilon''(\nu)/\nu$ spectra after EMT correction (Section 2.5).

***2.6.2. Damped harmonic oscillator model***

The BP feature in the normalized dielectric loss was fitted to a DHO profile:

$$(\varepsilon''(\nu))/(\nu)=(S\nu_0^2\Gamma\nu)/((\nu^2-\nu_0^2)^2+\Gamma^2\nu^2)+B(\nu)$$

where $\nu_0$ is the characteristic oscillator frequency, $\Gamma$ the damping coefficient (half-width), S the oscillator strength, and $B(\nu)$ a background function accounting for the non-BP continuum absorption. In practice, the low-frequency tail of $\varepsilon''(\nu)/\nu$ is dominated by a steep power-law decay unrelated to the BP. We therefore parameterize $B(\nu) = A\cdot\nu^{-p} + C$, where A, p, and C are fitted simultaneously with the DHO parameters. The three BP parameters, $\nu_0$, $\Gamma$, and S, were determined by nonlinear least-squares fitting over the 0.5–2.5 THz range. The oscillator strength S is proportional to the spectral weight of the BP and serves as a quantitative measure of the excess low-frequency dielectric absorption. The damping ratio $\Gamma/\nu_0$ characterizes the degree of inhomogeneous broadening of the excess vibrational modes.

### *2.6.3. Medium-range correlation length*

The BP frequency can be related to a structural correlation length through the scaling relation for disordered solids [15,16]:

$$\xi = (v_t)/(2\pi\ \nu_0)$$

where $v_t$ is the transverse (shear) wave velocity, estimated from the shear modulus G and density ρ of C-S-H as $v_t = (G/\rho)^{1/2}$. We adopt $G \approx 9$ GPa from molecular dynamics simulations of C-S-H [20] and use the measured pycnometric densities (Section 2.2). The length ξ represents a dynamical correlation length, the spatial scale over which low-frequency vibrational modes remain coherent, rather than the geometric size of any particular structural unit.

## 3. Results

### 3.1. Phase composition and purity

Fig. 1 shows the XRD patterns of the five C-S-H samples. All patterns exhibit the low-crystallinity, broad features characteristic of C-S-H gel: diffuse maxima near 2θ ≈ 29°, 32°, and 50° correspond to the average tobermorite-like silicate framework [36,37]. The low-angle region (5°–15°) shows a broad hump rather than sharp basal reflections, indicating turbostratic stacking disorder without long-range layer periodicity. No distinct diffraction peaks of $Ca(OH)_2$ (2θ ≈ 18°, 34°) or well-crystallized $CaCO_3$ (2θ ≈ 29.4°) are resolved, consistent with these phases being present only in minor amounts.

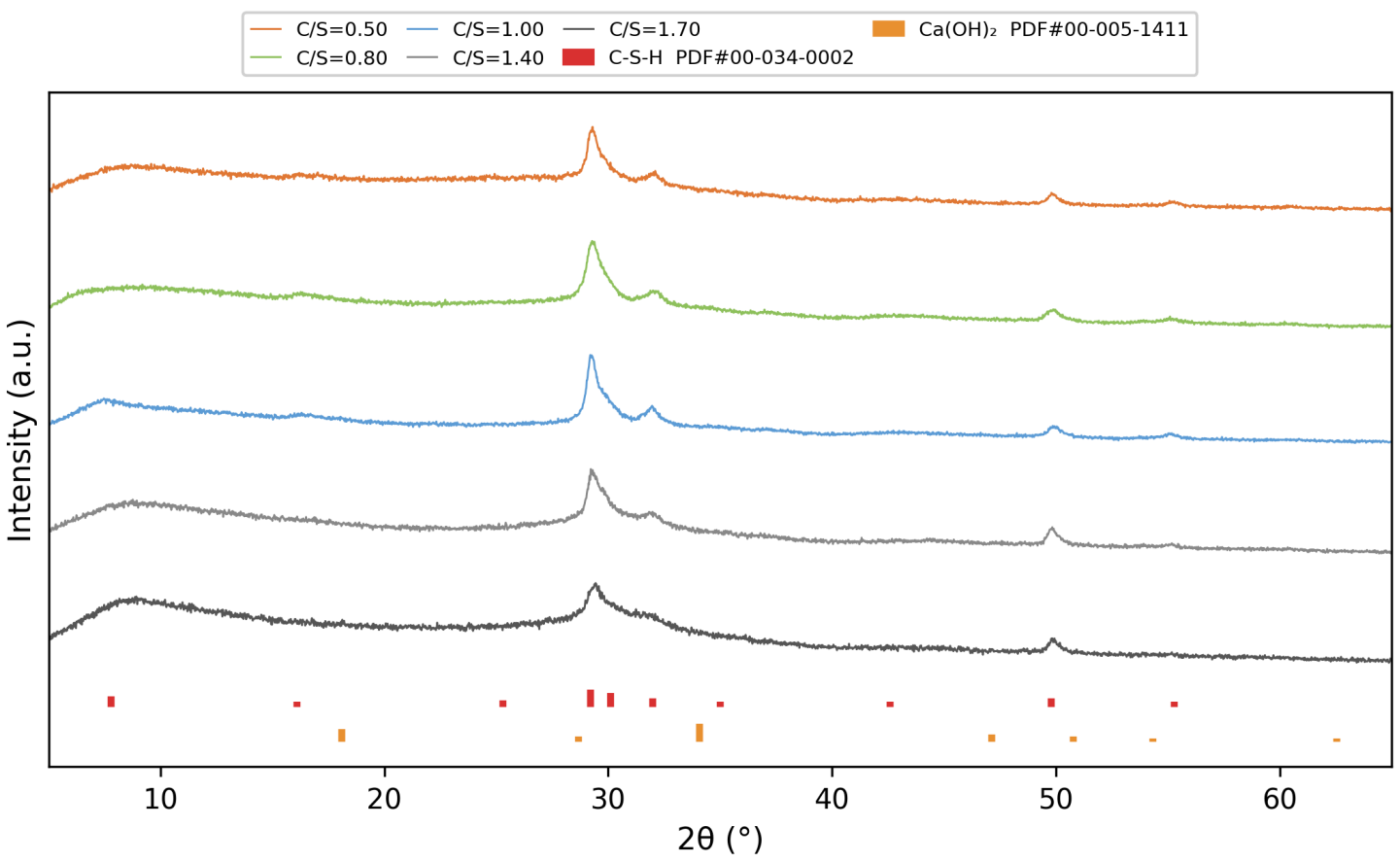

Fig. 1 XRD patterns of C-S-H at Ca/Si = 0.5–1.7, showing the broad, low-crystallinity features characteristic of C-S-H gel; reference stick patterns for C-S-H and $Ca(OH)_2$ are shown at the bottom

FTIR spectra (Fig. 2) confirm the expected C-S-H bonding environment. The strong absorption band at 900–1100 cm$^{-1}$, assigned to the asymmetric stretching vibration of Si-O in [$SiO_4$] tetrahedra [38], shifts slightly to lower wavenumber and broadens with increasing Ca/Si, reflecting progressive depolymerization of the silicate chains. The Si-O-Si bending mode near 450 cm$^{-1}$ and the symmetric bending near 660 cm$^{-1}$ (more prominent at low Ca/Si) further confirm the tobermorite-like layered framework [38,39]. Absorption bands in the 1400–1500 cm$^{-1}$ region indicate minor carbonate contamination. No absorption features attributable to residual organic solvents were detected in the 1700–1750 cm$^{-1}$ (C=O) or 2800–3000 cm$^{-1}$ (C-H) regions, confirming complete removal of isopropanol during drying.

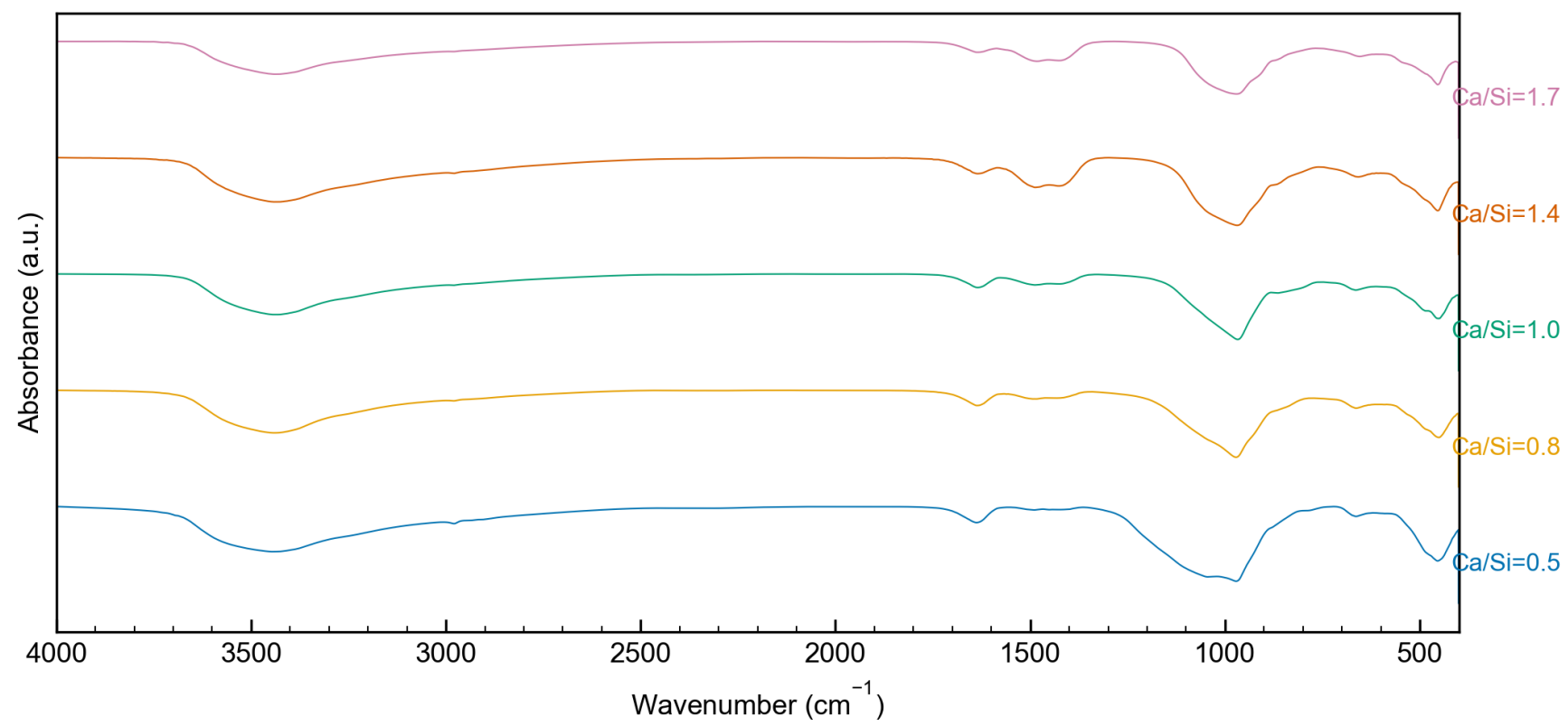

Fig. 2 FTIR spectra of C-S-H at Ca/Si = 0.5–1.7. The Si-O stretching band (900–1100 cm$^{-1}$) shifts to lower wavenumber and broadens with increasing Ca/Si

Table 1 summarizes the phase quantification derived from TGA. The weight loss was segmented into five temperature intervals: room temperature to ~105 °C (free and physically-adsorbed water), 105–200 °C (interlayer water), 200–350 °C (structural water and further C-S-H dehydration), 400–500 °C ($Ca(OH)_2$ dehydroxylation), and 600–800 °C ($CaCO_3$ decomposition). The $Ca(OH)_2$ content ranges from 4.3% (Ca/Si = 0.8) to 9.2% (Ca/Si = 1.7), increasing with Ca/Si as expected from the greater excess of calcium at high ratios. $CaCO_3$ remains below 2% for all samples, consistent with the absence of sharp calcite reflections in XRD. The structural water fraction (200–350 °C) increases from 2.7% at Ca/Si = 0.5 to 4.1% at Ca/Si = 1.7, with an inflection near Ca/Si = 1.0 (3.4%) that signals a change in the binding environment of interlayer water, a point we revisit in the Discussion. The actual Ca/Si ratios of the C-S-H phase, computed from XRF data after subtracting calcium in $Ca(OH)_2$ and $CaCO_3$, are 0.54, 0.90, 1.14, 1.46, and 1.57 for nominal values of 0.5, 0.8, 1.0, 1.4, and 1.7, respectively.

Table 1 Composition of C-S-H samples determined by TGA and XRF

| Nominal Ca/Si | Free water (%) | Interlayer water (%) | Structural water (%) | $Ca(OH)_2$ (%) | $CaCO_3$ (%) | Actual Ca/Si |
|---|---|---|---|---|---|---|
| 0.5 | 6.45 | 6.59 | 2.76 | 7.05 | 1.31 | 0.54 |
| 0.8 | 7.30 | 7.65 | 2.72 | 4.33 | 0.04 | 0.90 |
| 1.0 | 7.83 | 6.36 | 3.44 | 5.11 | 1.70 | 1.14 |
| 1.4 | 6.51 | 6.40 | 3.43 | 6.84 | 1.81 | 1.46 |
| 1.7 | 5.45 | 5.67 | 4.10 | 9.17 | 1.72 | 1.57 |

Laser diffraction shows that the C-S-H particles are predominantly in the 3–30 μm range with a maximum below 100 μm. Since the THz wavelength at 3 THz (~100 μm) exceeds the largest particle dimension, scattering losses in the BP frequency region (0.4–1.5 THz, $\lambda > 200$ μm) are in the Rayleigh regime and negligible. The measured true (skeletal) densities are 2.116, 2.165, 2.323, 2.279, and 2.768 g/cm$^3$ for Ca/Si = 0.5 through 1.7, respectively. The density increases overall with Ca/Si, as heavier CaO replaces lighter $SiO_2$ in the structure, but exhibits a local maximum at Ca/Si = 1.0 (2.323 g/cm$^3$) followed by a slight dip at Ca/Si = 1.4 (2.279 g/cm$^3$) before rising sharply at Ca/Si = 1.7. This non-monotonicity at intermediate Ca/Si echoes the structural crossover discussed later and may reflect a change in packing efficiency as the silicate chain topology reorganizes.

### 3.2. Intrinsic THz dielectric response

Fig. 3a displays representative time-domain waveforms after transmission through the C-S-H/COC pellets, together with the air reference. All sample waveforms show reduced amplitude and a time delay relative to the reference, indicating absorption and phase retardation by the C-S-H composite. Fig. 3b presents the corresponding power spectra; signal quality remains adequate over 0.4–3.0 THz for all samples.

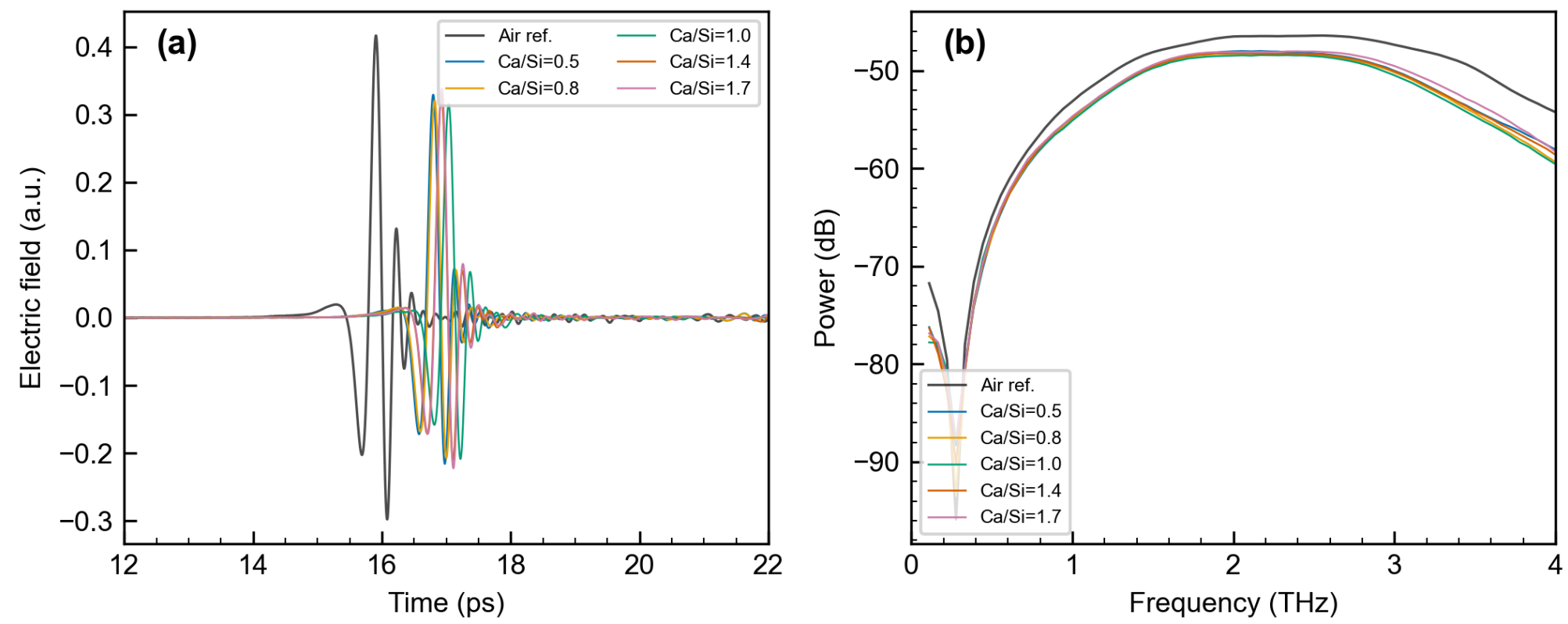

Fig. 3 (a) THz time-domain waveforms and (b) power spectra for the five C-S-H/COC pellets and the air reference

After extracting the optical constants of the composite pellets and applying the Bruggeman EMT correction, we obtain the intrinsic dielectric properties of C-S-H (Fig. 4). The refractive index (Fig. 4a) is nearly frequency-independent above 0.5 THz, a weak-dispersion profile typical of ionic/covalent solids in the far-infrared. This flatness contrasts sharply with the strong anomalous dispersion of liquid water in the same frequency range, where the Debye relaxation tail produces a steep decrease in n(ν). The absence of such a feature in our spectra confirms that the THz response arises from the solid silicate-calcium framework rather than from residual free water. The mean refractive index (averaged over 0.5–2.5 THz) varies non-monotonically with Ca/Si: 1.59 (Ca/Si = 0.5), 1.68 (0.8), 1.85 (1.0), 1.75 (1.4), and 1.67 (1.7), rising to a maximum at Ca/Si = 1.0 and decreasing thereafter. The real permittivity ε' (Fig. 4b) follows the same trend, as expected from $\varepsilon' \approx n^2$ at low loss, reaching a maximum of ~3.4 at Ca/Si = 1.0. As a composite property influenced by the solid framework, structural water, and residual porosity, the refractive index and permittivity are not straightforward to interpret mechanistically; we note the extremum at Ca/Si = 1.0 as a further indicator of a structural change at this composition without attributing it to a single cause.

The absorption coefficient (Fig. 4c) increases with frequency following a power-law envelope $\alpha \propto \nu^s$. Fitting over 1.5–2.5 THz (above the BP region) yields $s \approx 1$ for Ca/Si = 0.5 (at the lower compositional boundary) and somewhat above unity, up to ~1.5, for the higher ratios. These values lie between the linear dependence (s=1) associated with coupling to a flat density of states and the quadratic frequency dependence (s=2) expected for Rayleigh scattering of phonons by point-like disorder, consistent with the behavior reported for other disordered solids in this frequency range [40]. At 1 THz, the intrinsic absorption coefficient ($\alpha = 4\pi\nu\kappa/c$, with $\kappa = \varepsilon''/2n$) ranges from approximately 6 $cm^{-1}$ (Ca/Si = 1.7) to 9 $cm^{-1}$ (Ca/Si = 1.0).

More revealing is the dielectric loss ε" (Fig. 4d): all five samples exhibit a broad absorption feature centered near 1 THz, superimposed on the rising power-law background. This feature, absent in the reference spectra of crystalline $Ca(OH)_2$ and $CaCO_3$ measured under identical conditions, is the spectroscopic signature we identify as the boson peak.

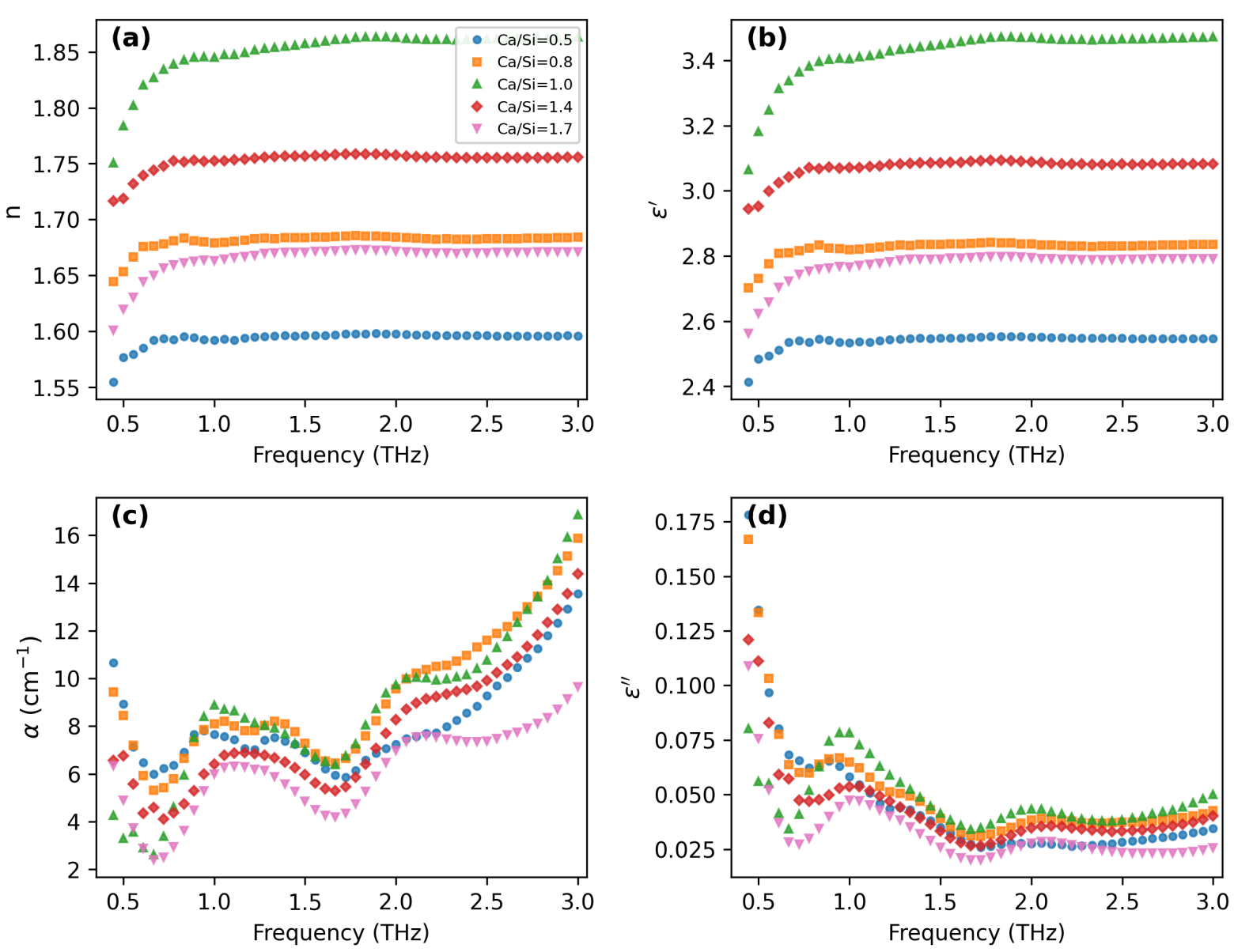

Fig. 4 Intrinsic THz optical properties of C-S-H after Bruggeman effective-medium correction: (a) refractive index n, (b) real permittivity ε′, (c) absorption coefficient α, and (d) dielectric loss ε″, for Ca/Si = 0.5–1.7

### 3.3. THz boson peak: identification and DHO fitting

Fig. 5 shows the frequency-normalized dielectric loss $\varepsilon''(\nu)/\nu$ for all five Ca/Si ratios. The normalization suppresses the trivial ν-linear trend and makes the BP emerge as a distinct broad hump in the 0.7–1.2 THz range. The peak is present in every sample, ruling out an instrumental or sample-preparation artifact. Each composition was measured more than ten times by THz-TDS, and the boson peak and its compositional trend were reproducible throughout; the calorimetric $C_p/T^3$ excess was likewise reproduced on an independently prepared Ca/Si = 1.7 specimen (Section 2.4).

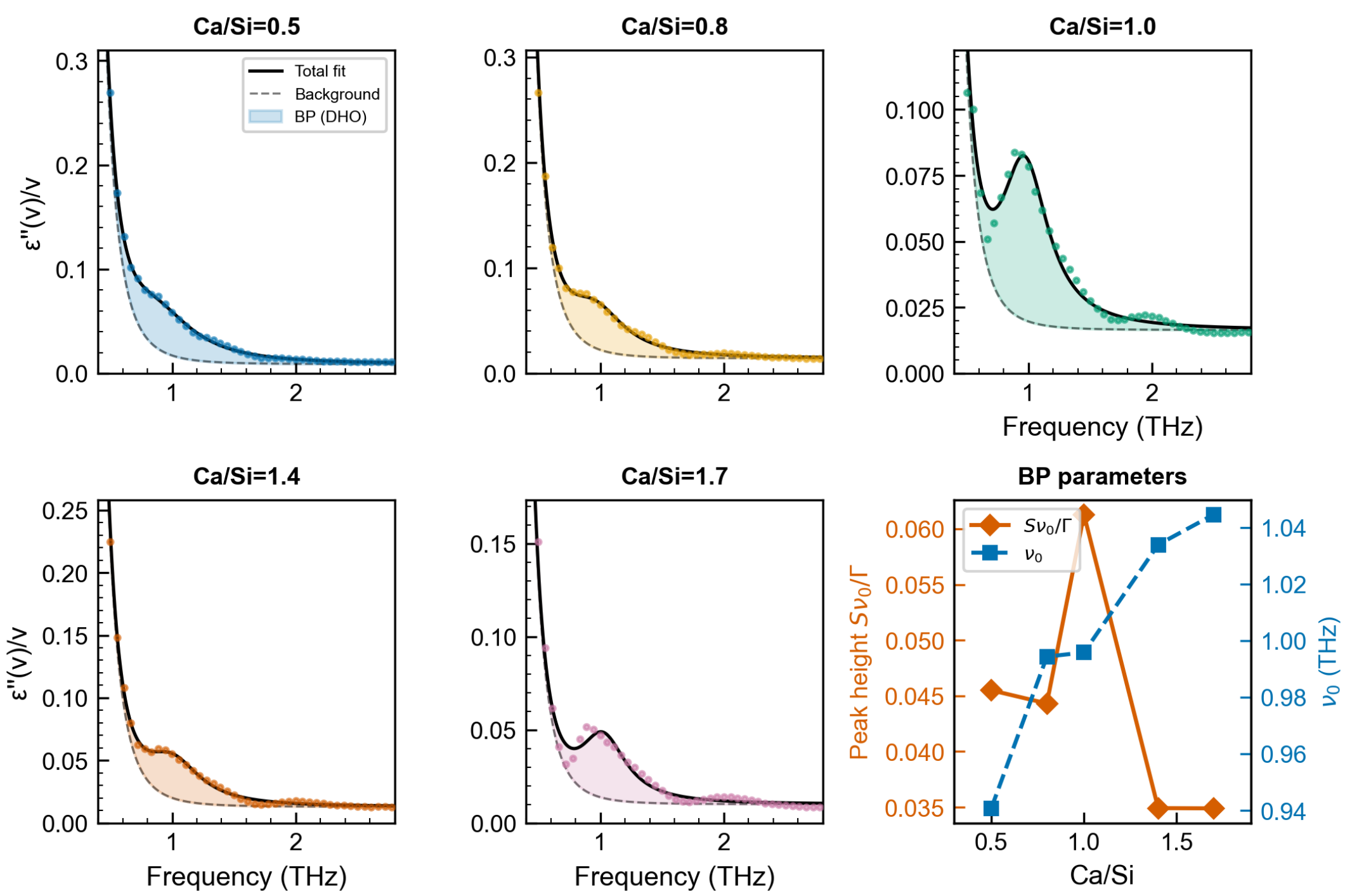

Fig. 5 Frequency-normalized dielectric loss ε″(ν)/ν (symbols) and damped-harmonic-oscillator fits (lines), revealing the boson peak near 1 THz

Because the low-frequency portion of ε''(ν)/ν is dominated by a steep relaxation-type tail unrelated to the BP, the spectra were fitted to a composite model consisting of a power-law background $A \cdot \nu^{-p} + C$ and a DHO peak (Section 2.6.2). Solid curves in Fig. 5 show the resulting least-squares fits over the 0.5–2.5 THz range; shaded regions indicate the DHO contribution alone. The fits reproduce the data well for all five samples ($R^2$ = 0.966–0.999). The fitted parameters are collected in Table 2.

Table 2 DHO fitting parameters for the THz boson peak of C-S-H

| Ca/Si | $\nu_0$ (THz) | Γ (THz) | $\Gamma/\nu_0$ | S | $S \cdot \nu_0/\Gamma$ | $R^2$ |
|---|---|---|---|---|---|---|
| 0.5 | 0.94 | 0.79 | 0.84 | 0.038 | 0.046 | 0.999 |
| 0.8 | 0.99 | 0.61 | 0.62 | 0.027 | 0.044 | 0.996 |
| 1.0 | 1.00 | 0.47 | 0.48 | 0.029 | 0.061 | 0.978 |
| 1.4 | 1.03 | 0.58 | 0.56 | 0.020 | 0.035 | 0.996 |
| 1.7 | 1.05 | 0.47 | 0.45 | 0.016 | 0.035 | 0.966 |

The column $S \cdot \nu_0/\Gamma$ gives the DHO peak height above the background, the maximum of the DHO function at $\nu \approx \nu_0$, and represents the visible amplitude of the BP in the spectrum. Three features of this table deserve attention.

First, the characteristic frequency $\nu_0$ increases mildly from 0.94 THz at Ca/Si = 0.5 to 1.05 THz at Ca/Si = 1.7, but the variation is modest (~11%) compared to the factor-of-two changes in other parameters. The values fall in the same low-THz band as the MD prediction of Abdolhosseini Qomi et al. [20] and the BP frequencies of silicate glasses [17,18], although they lie systematically below the MD estimate (Section 4.6).

Second, the damping ratio $\Gamma/\nu_0$ is sub-unity for all samples (0.45–0.84), indicating that the BP modes are underdamped but substantially broadened. $\Gamma/\nu_0$ follows a non-monotonic pattern, reaching a local minimum at Ca/Si = 1.0 (0.48) and its lowest value at Ca/Si = 1.7 (0.45); the BP modes at these compositions are the most coherent; their vibrational character is least disrupted by anharmonic decay or scattering.

Third, and most revealing, the oscillator strength S and the peak height $S \cdot \nu_0/\Gamma$ tell different stories. S decreases roughly monotonically from 0.038 (Ca/Si = 0.5) to 0.016 (Ca/Si = 1.7), tracking the same trend as the calorimetric $C_p/T^3$ peak height (Section 3.4). The peak height, however, reaches a clear maximum at Ca/Si = 1.0 (0.061, nearly twice the value at Ca/Si = 1.7). This decoupling arises because the damping $\Gamma$ drops sharply at Ca/Si = 1.0: even though the total spectral weight S is not the largest, the narrower peak concentrates the available oscillator strength into a taller, more prominent feature. In physical terms, Ca/Si = 1.0 does not have the most excess vibrational modes, but those modes are the least damped and the most dipole-active per unit spectral weight. This non-monotonic peak-height trend is robust to the impurity correction (Section 2.5), confirming that it is intrinsic to C-S-H.

### 3.4. Low-temperature heat capacity

The measured heat capacity $C_p(T)$ rises monotonically with temperature for all five samples from 2 to 300 K. At 300 K the values cluster between 0.97 and 1.04 J/(g·K), consistent with the Dulong-Petit limit for a material with an effective atomic mass around 20–25 g/mol and comparable to values reported for C-S-H by Bentz and others [41].

To separate the various contributions to the low-temperature heat capacity, we fit the data below 5 K to the standard two-term expression $C_p = \gamma T + \beta T^3$, where $\gamma T$ represents the two-level system (TLS) contribution characteristic of amorphous solids and $\beta T^3$ is the Debye phonon term. The fitted coefficients are listed in Table 3.

Table 3 Low-temperature heat capacity parameters from $C_p = \gamma T + \beta T^3$ fit (T < 5 K)

| Ca/Si | $\gamma$ ($10^{-6}$ J $g^{-1}$ $K^{-2}$) | $\beta$ ($10^{-6}$ J $g^{-1}$ $K^{-4}$) | $\Theta_D$ from $\beta$ (K) |
|---|---|---|---|
| 0.5 | 5.19 | 3.50 | ~240 |
| 0.8 | 10.35 | 2.58 | ~265 |
| 1.0 | 8.64 | 2.39 | ~272 |
| 1.4 | 7.00 | 2.25 | ~278 |
| 1.7 | 10.19 | 2.06 | ~285 |

The linear coefficient $\gamma$ is non-zero for all samples (5–10 × $10^{-6}$ J $g^{-1}$ $K^{-2}$), consistent with the glassy character of C-S-H: crystalline solids have $\gamma = 0$ (no electronic or TLS contribution), while amorphous materials universally exhibit a finite linear term arising from tunneling between nearly degenerate local energy minima [42]. In our hydrated, powdered samples this two-level-system population may include contributions from the confined interlayer water and from surface states in addition to the silicate framework, which the calorimetry cannot separate. The Debye coefficient $\beta$ generally decreases from Ca/Si = 0.5 to 1.7, corresponding to an increase in the Debye temperature from ~240 to ~285 K, with a slight non-monotonicity at Ca/Si = 1.4. These calorimetric Debye temperatures are somewhat lower than those estimated from sound velocity in Section 4.4 (~290–300 K), likely because the Debye $T^3$ law holds strictly only well below $\Theta_D/50$ and our fitting range (2–5 K) may still contain residual BP contributions that inflate $\beta$. This $\beta$-derived $\Theta_D$ moreover rises with Ca/Si, opposite to the sound-velocity estimate (Section 4.4); the two are reconciled by the same residual BP excess, which is largest at low Ca/Si (Fig. 6a) and there suppresses the apparent $\Theta_D$ most, so the $\beta$-derived values reflect BP contamination rather than the true Debye temperature (taken from sound velocity).

The more informative representation is $C_p/T^3$ versus T (Fig. 6a). Every sample shows a broad hump well above the flat Debye baseline $\beta$, peaking between 7 and 12 K (6.8–11.7 K after correction). This excess is the thermodynamic manifestation of the BP, an unambiguous sign that C-S-H possesses more low-frequency vibrational states than predicted by the Debye model.

After correcting for the crystalline $Ca(OH)_2$ and $CaCO_3$ contributions (Section 2.5), the peak heights follow the order Ca/Si = 0.5 ($4.14 \times 10^{-6}$ J/(g·K$^4$)) > 0.8 ($3.06 \times 10^{-6}$) > 1.0 ($2.80 \times 10^{-6}$) > 1.7 ($2.74 \times 10^{-6}$) ≈ 1.4 ($2.68 \times 10^{-6}$). The correction changed the peak heights by less than 5% and did not alter the compositional trend (Fig. A.1). Unlike the THz BP peak height, the calorimetric peak intensity decreases roughly monotonically with increasing Ca/Si, with no maximum at Ca/Si = 1.0. The peak temperatures are 9.9, 7.9, 9.4, 11.7, and 6.8 K for Ca/Si = 0.5, 0.8, 1.0, 1.4, and 1.7, respectively. The scatter in peak temperature (6.8–11.7 K) reflects the broad, flat-topped shape of the $C_p/T^3$ hump, which makes the apparent maximum sensitive to noise; the energy scale is nonetheless consistent across all samples. We return to the comparison between THz and calorimetric BP intensities in Section 4.2.

These analyses are framed within the soft potential model (SPM), in which the two-level tunnelling systems (the $\gamma$ T term), the boson peak (soft quasi-localised modes), and the far-infrared/THz absorption all arise from a single distribution of soft anharmonic potentials obeying the universal $g(\omega) \sim \omega^4$ law [43,44]. To test whether the BP frequency can be read directly from $C_p$, we fitted the low-temperature heat capacity to five models of increasing physical content (Appendix A.1). All reproduce $C_p(T)$ comparably well yet return markedly different BP frequencies, because the $C_p(T)$ integral is dominated by the total excess mode count rather than the frequency distribution: the Einstein-oscillator models are non-unique (Einstein temperatures differing by up to a factor of two at indistinguishable fit quality) and a regularised VDOS inversion shows no resolvable peak.

Only the physically motivated model (a Planck-like parametrisation of the SPM soft-mode VDOS, $g_q(\nu)/\nu^2 = \varphi \, \nu^3/(e^{\ell\nu}-1)$, whose $\nu^4$ low-frequency limit matches the universal scaling of quasi-localised modes [45]) yields a well-defined frequency. A four-parameter fit over 2–30 K returns $\nu_{BP}$ = 1.13–1.34 THz, agreeing with the THz DHO $\nu_0$ to within ~20% (the residual offset is expected, as $g_q(\nu)/\nu^2$ is the bare reduced VDOS whereas the THz $\varepsilon''(\nu)/\nu$ peak is shifted by the frequency-dependent dipole coupling); $\nu_{BP}$ is robust to the fit cutoff (<5% from 30 to 50 K), and only 0.8–1.7% of modes are quasi-localised, consistent with the BP being a minor feature on a dominant Debye background. The per-composition dual-axis decomposition is shown in Fig. 6b–f. The heat capacity does not uniquely select the Planck form (any asymmetric VDOS sharing the $\omega^4$ limit fits comparably), so the choice rests on physical grounds and the absolute frequency is fixed by THz. For C-S-H, an insulator, the γ T term represents the two-level-system (tunnelling) contribution rather than the electronic term of the original metallic-glass formulation, a clear fingerprint of the glassy soft-potential landscape.

The most stringent cross-check is parameter-free (Appendix A.1): transferring the spectroscopically determined DHO spectral function directly into a forward prediction of $C_p/T^3$, with no adjustable BP frequency or width, reproduces the measured calorimetric excess for four of the five compositions, the exception being the extremely broad Ca/Si = 0.5 profile (Fig. A.3), confirming quantitative self-consistency between the two independent measurements. The broader lesson is that the standard calorimetric models, all developed for melt-quenched glasses, reproduce the C-S-H heat capacity yet blur the BP frequency, precisely because the C-S-H BP is far broader and more heavily damped ($\Gamma/\nu_0$ = 0.45–0.84) than those of conventional glasses ($\Gamma/\nu_0 \sim$ 0.2–0.3): for heavily damped BPs, direct spectroscopic detection (THz-TDS) is essential, while calorimetry validates the result through forward modelling.

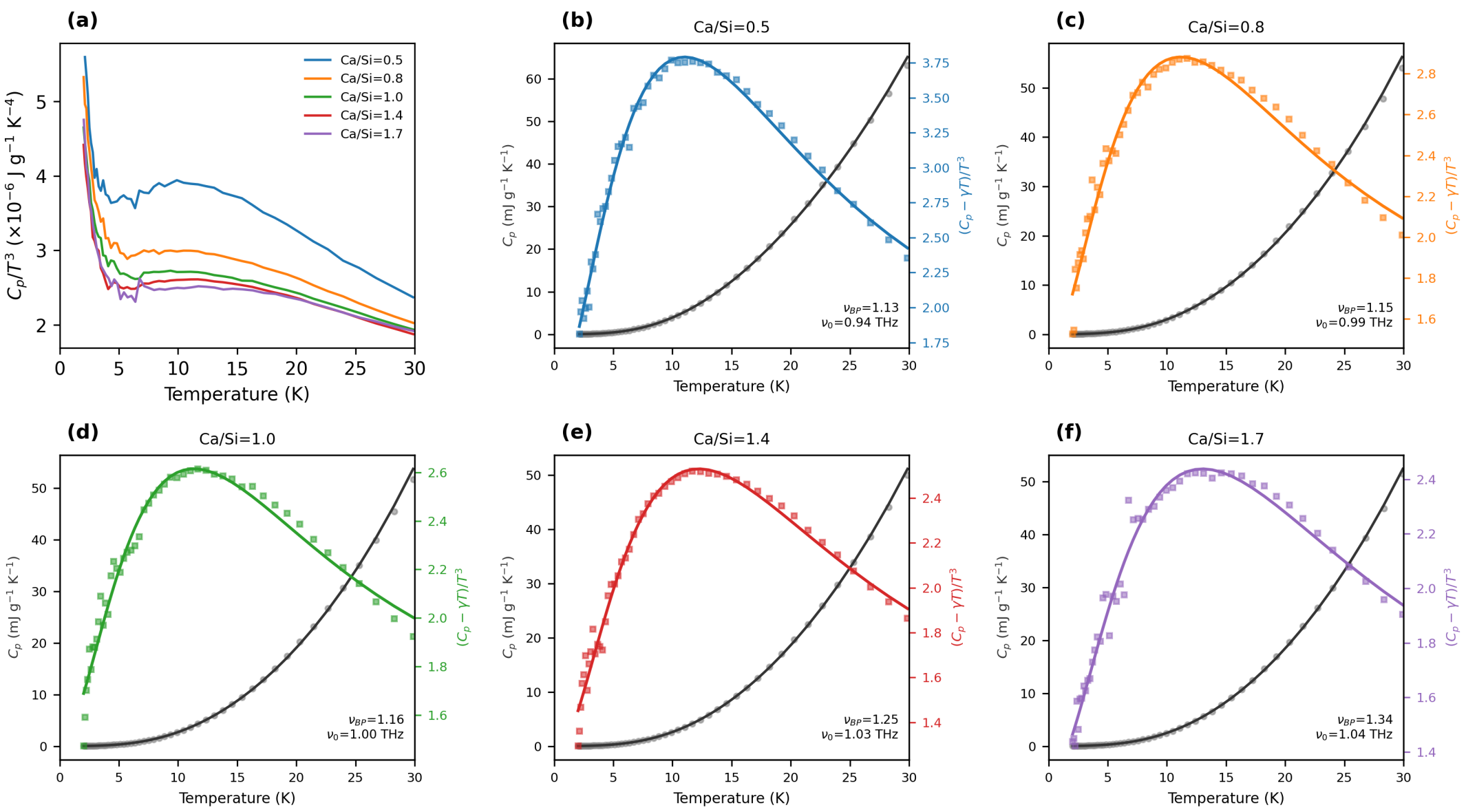

Fig. 6 Low-temperature heat capacity and Planck soft-mode analysis. (a) $C_p/T^3$ versus T for all five Ca/Si, showing the boson-peak excess hump at 7–12 K. (b–f) Per-composition Planck-like soft-mode fits over the 2–30 K window (dual-axis after Ding et al.: left axis $C_p$, right axis $(C_p-\gamma T)/T^3$; symbols, data; lines, fit), each annotated with the calorimetric $\nu_B$P and the THz $\nu_0$

## 4. Discussion

### 4.1. Cross-validation of the boson peak: frequency correspondence

The two independent probes, THz-TDS and low-temperature calorimetry, detect the BP through fundamentally different physical mechanisms, yet both place the characteristic excitation energy in the same frequency range. To make the comparison quantitative, we convert the $C_p/T^3$ peak temperatures $T_{BP}$ to equivalent frequencies using $f = k_B T_{BP}/h$. This direct conversion yields 0.14–0.24 THz, substantially below the ~1 THz THz-TDS peak. The offset is expected and well documented in the glass literature: $C_p/T^3$ is an integral quantity weighted by the Bose-Einstein occupation factor, which amplifies the contribution of the lowest-frequency modes and shifts the apparent peak to temperatures (and hence equivalent frequencies) below the true $g(\omega)/\omega^2$ maximum [46,47]. In binary $SiO_2$–$Al_2O_3$ glasses, Ando et al. [46] compared BP frequencies obtained simultaneously by Raman, THz-TDS, and calorimetry and found the calorimetric peak temperature systematically lower by a factor of 4–5. Applying the same empirical correction, the calorimetric frequencies fall in the 0.6–1.2 THz range (Table 4), in reasonable agreement with the THz-TDS values.

Table 4 Comparison of BP frequencies from THz-TDS and low-temperature heat capacity

| Ca/Si | $C_p/T^3$ peak T (K) | $k_B$ T/h (THz) | Corrected (×4–5) (THz) | THz DHO $\nu_0$ (THz) |
|---|---|---|---|---|
| 0.5 | 9.9 | 0.21 | 0.83–1.03 | 0.94 |
| 0.8 | 7.9 | 0.17 | 0.66–0.83 | 0.99 |
| 1.0 | 9.4 | 0.20 | 0.78–0.98 | 1.00 |
| 1.4 | 11.7 | 0.24 | 0.97–1.22 | 1.03 |
| 1.7 | 6.8 | 0.14 | 0.57–0.71 | 1.05 |

The agreement, after accounting for the known systematic bias of calorimetry, constitutes strong independent confirmation that the broad spectral feature observed by THz-TDS is indeed the BP, and not a measurement artifact or an impurity-related absorption. The shift factor depends on the BP width (broader peaks shift more), and our individual ratios $\nu_0/(k_B T_{BP}/h)$ range from 4.3 (Ca/Si = 1.4) to 7.5 (Ca/Si = 1.7), with a median near 5, consistent with values reported for silicate (~4.5) and lithium borate (6–7) glasses [46,48]; the somewhat larger scatter reflects the broad, flat-topped $C_p/T^3$ humps, whose apparent peak temperature is sensitive to noise.

**4.2. Elastic heterogeneity and the intensity decoupling**

The most informative aspect of combining THz-TDS with calorimetry is not the agreement in peak position, that merely confirms both probes see the same phenomenon, but the disagreement in peak intensity, which the DHO decomposition now allows us to dissect quantitatively.

Three quantities track the BP "intensity" and follow distinct trends with Ca/Si (Fig. 7; Table 2). Both the calorimetric $C_p/T^3$ peak height and the DHO oscillator strength S, each a measure of the *total* excess mode count (the former over all modes, the latter over the dipole-active ones), decrease roughly monotonically from Ca/Si = 0.5 to 1.7, so low-Ca/Si C-S-H, with its long, highly connected silicate chains, possesses the most excess vibrational modes. The DHO peak height $S \cdot \nu_0/\Gamma$, by contrast, peaks sharply at Ca/Si = 1.0 (0.061, nearly twice the value at 1.4 or 1.7): there the damping is low ($\Gamma = 0.47$ THz) while S remains substantial, concentrating the spectral weight into a narrow, tall peak, whereas Ca/Si = 1.7 has equally low damping but less than half the weight. The apparent "strongest BP at Ca/Si = 1.0" that dominates visual inspection of the $\varepsilon''(\nu)/\nu$ spectra therefore reflects not the most excess modes but the most coherent, dipole-active ones.

The suppression of damping at Ca/Si ≈ 1.0 reflects the structural crossover analysed in Section 4.5: with the bridging silicate tetrahedra largely depleted but interlayer calcium not yet filling, the framework is least constrained, so the low-frequency modes encounter fewer scattering centers (broken chains, calcium clusters) that would broaden the BP, while the disrupted, charge-asymmetric network couples them efficiently to the THz field. The result is a tall, relatively narrow BP at Ca/Si ≈ 1.0 even as the total excess mode count (S and $C_p/T^3$) continues to fall.

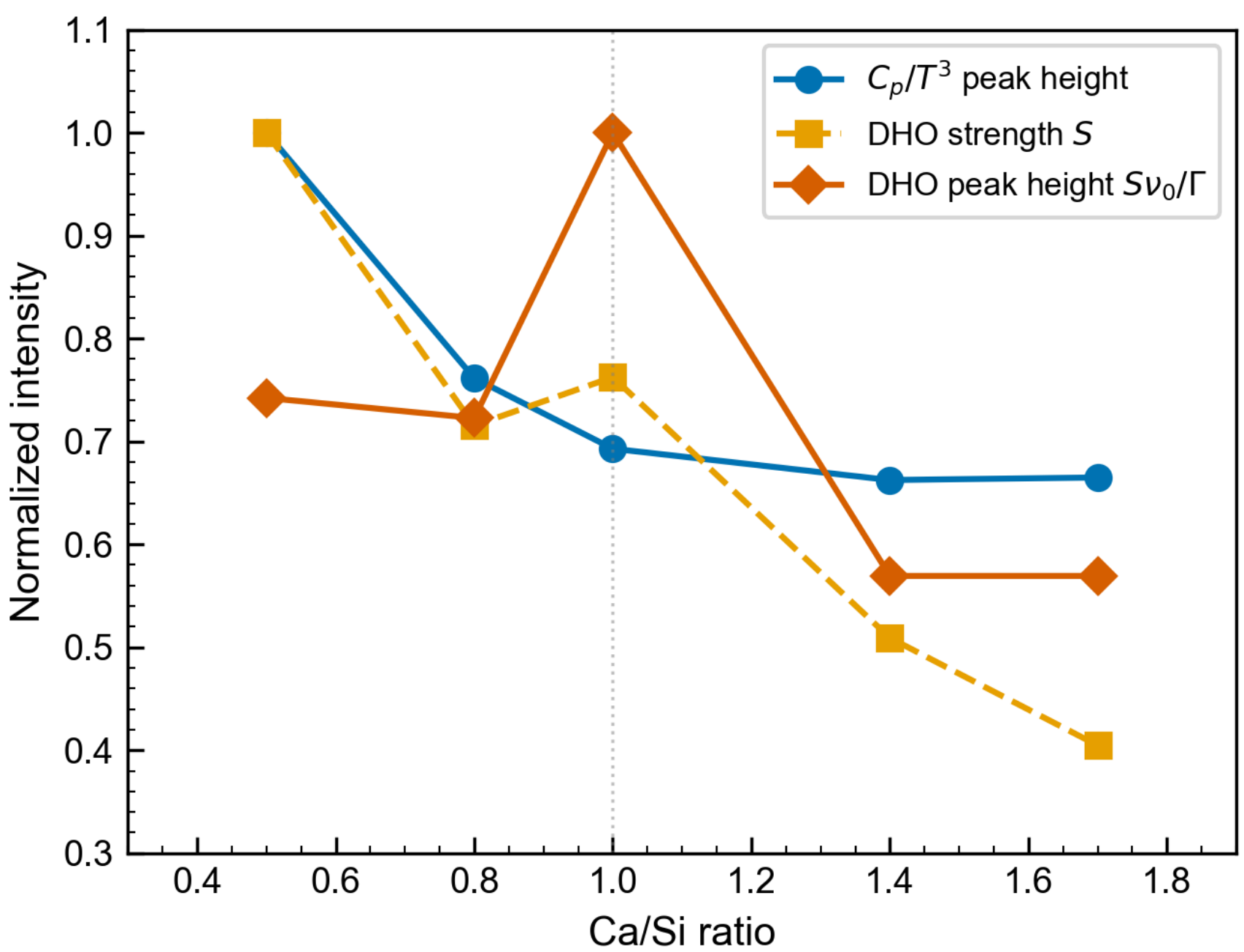

Fig. 7 Three boson-peak intensity indicators versus Ca/Si: integrated DHO strength S, THz peak height $S \cdot \nu_0/\Gamma$, and calorimetric $C_p/T^3$ peak height, each normalized to its maximum. The dotted line marks the Ca/Si ≈ 1.0 crossover

More broadly, the apparent BP "intensity" in a THz spectrum is not a simple proxy for the total excess density of states, nor for the total structural disorder. It is a compound quantity shaped by the mode count, the dipole coupling strength, and the damping. Disentangling these contributions requires either the kind of DHO decomposition performed here, or an independent measurement (such as calorimetry) that is insensitive to dipole selection rules. The phenomenological DHO used in Section 3.3 can be placed on firmer theoretical ground through the spatially heterogeneous elasticity (SHE) framework of Schirmacher et al. [49,50]. In a medium whose shear modulus fluctuates spatially, as expected in turbostratically disordered C-S-H, the self-consistent Born approximation (SCBA) applied to the elastic wave equation yields an excess VDOS that takes a DHO form near the Ioffe-Regel crossover. The key parameter is $\gamma$, the dimensionless variance of the shear modulus fluctuation field, which simultaneously controls both the BP amplitude and the phonon scattering rate. Unlike the phenomenological DHO parameters of Section 3.3, $\gamma$ is therefore a direct physical measure of the medium-range elastic heterogeneity of C-S-H. The resulting multi-physics superposition formula for the normalized dielectric loss is [49,50]:

$$(\varepsilon''(\nu))/(\nu) = (A_{bg})/(\nu^s) + (C_{BP}\cdot\gamma)/([1-((\nu)/(\nu_{BP}))^2]^2 + ((\gamma\cdot\nu)/(\nu_{BP}^2))^2) + C_D$$

The three terms have distinct physical origins: (i) $A_{bg}/\nu^s$ is the high-frequency tail of Cole-Cole relaxation from nanoconfined pore and interlayer water, whose relaxation peak lies below the measurement window; (ii) the CPA-derived DHO describes the boson peak arising from phonon scattering by elastic heterogeneity, with $\gamma$ encoding the degree of medium-range elastic disorder; and (iii) $C_D$ is the Debye phonon dissipation baseline. Unlike the standard DHO (Section 2.6.2), this formulation does not treat oscillator strength and damping as independent parameters; both are governed by $\gamma$ alone.

We fitted this formula to the same $\varepsilon''(\nu)/\nu$ data over 0.5–2.5 THz. The results (Table 5, Fig. A.4) are in close agreement with the phenomenological DHO analysis.

Table 5 CPA multi-physics superposition fit parameters

| Ca/Si | $A_{bg}$ | s | $C_{BP}$ | $\gamma$ | $\nu_{BP}$ (THz) | $C_D$ | $R^2$ |
|---|---|---|---|---|---|---|---|
| 0.5 | 0.0037 | 5.82 | 0.041 | 0.98 ± 0.08 | 1.02 ± 0.02 | 0.010 | 0.998 |
| 0.8 | 0.0040 | 6.00 | 0.028 | 0.70 ± 0.07 | 1.05 ± 0.01 | 0.015 | 0.996 |
| 1.0 | 0.0007 | 6.00 | 0.030 | 0.54 ± 0.06 | 1.03 ± 0.01 | 0.018 | 0.980 |
| 1.4 | 0.0049 | 5.40 | 0.018 | 0.63 ± 0.06 | 1.09 ± 0.01 | 0.014 | 0.996 |
| 1.7 | 0.0020 | 6.00 | 0.014 | 0.48 ± 0.07 | 1.08 ± 0.02 | 0.011 | 0.967 |

First, the CPA disorder parameter $\gamma$ decreases from 0.98 (Ca/Si = 0.5) to 0.48 (Ca/Si = 1.7), confirming that the elastic heterogeneity decreases systematically as the structure transitions from a long-chain, highly polymerized network to a calcium-rich, partially stiffened phase. Second, $\gamma$ tracks the phenomenological damping ratio $\Gamma/\nu_0$ from the standard DHO fit (Table 2) to within ~15%, validating the physical correspondence between the two parameterizations. Third, the BP frequency $\nu_{BP}$ from the CPA fit (1.02–1.09 THz) agrees closely with the standard DHO $\nu_0$ (0.94–1.05 THz), with the CPA values systematically ~5% higher, a minor shift attributable to the different functional forms of the two DHO variants.

The water relaxation term $A_{bg}/\nu^s$ has an exponent s = 5.4–6.0, much steeper than the Debye limit (s = 1) or the Cole–Cole limit (s < 2), indicating that the relaxation peak of the confined water lies far below 0.5 THz and contributes to the measured window only as a rapidly decaying tail. This reflects the tightly-bound, strongly-confined nature of the interlayer water retained in our partially-dried samples (~6–8 wt%, Table 1), whose dielectric response is suppressed and anisotropic [51,52]; despite its substantial abundance, this confined water contributes negligibly above 0.7 THz. Consistent with a framework origin, the BP parameters ($\nu_0$, S, Γ and γ) track Ca/Si far more strongly than the water content: they vary systematically with composition while the total and interlayer water (Table 1) do not, so the boson peak is governed by the silicate framework rather than by the confined water. These values carry an important caveat: because the BP analysis is performed on a powder, the extracted γ, the velocities, and the correlation length of Section 4.3 are orientation-averaged effective quantities. C-S-H is intrinsically anisotropic, its stiff silicate sheets and soft interlayer giving a larger modulus contrast, and a correspondingly shorter correlation length, across the sheets than within them, so the true elastic heterogeneity is directional and a powder measurement cannot resolve it. The compositional trend in γ is nonetheless robust, since all samples are measured identically; resolving the anisotropy would require oriented specimens or wavevector-resolved inelastic neutron scattering.

**4.3. The medium-range dynamical correlation length**

The boson-peak frequency marks a well-defined dynamical crossover. Numerical and theoretical work has established that $\nu_0$ coincides with the Ioffe-Regel limit for transverse phonons, the frequency above which transverse waves cease to propagate because their mean free path has fallen to the order of their wavelength [53,54]. Within the spatially-heterogeneous-elasticity picture this is the same crossover at which the medium-range shear-modulus fluctuations (the parameter γ of Section 4.2) come to dominate the dynamics [49,50]. The boson peak therefore encodes a structural length, but mapping the frequency onto a real-space value requires a prefactor that is not unique.

Three conventions bracket this length (Table 6). The reduced-wavelength (plane-wave) relation $\xi_1 = v_t/(2\pi\nu_0)$ treats $\nu_0$ as a sharp Ioffe-Regel cutoff and gives $\xi_1 \approx 0.27$–$0.35$ nm. The confined-mode relation $\xi_2 = S_0\, v_t/(2\nu_0)$ ($S_0 \approx 0.7$), introduced by Duval et al. [16] for the eigenmodes of nanoscale elastic domains, gives $\xi_2 \approx 0.60$–$0.77$ nm. The Ioffe-Regel wavelength itself, $\lambda_{IR} = v_t/\nu_0$, the actual scale below which transverse propagation ceases, gives $\lambda_{IR} \approx 1.7$–$2.2$ nm. Throughout, $v_t = (G/\rho)^{1/2} \approx 1800$–$2060$ m/s from $G \approx 9$ GPa [20] and the measured pycnometric densities.

Table 6 Length scales associated with the boson peak under three conventions (order-of-magnitude estimates)

| Ca/Si | $\nu_0$ (THz) | $v_t$ (m/s) | $\xi_1 = v_t/(2\pi\nu_0)$ (nm) | $\xi_2 = 0.7\ v_t/(2\nu_0)$ (nm) | $\lambda_{IR} = v_t/\nu_0$ (nm) |
|---|---|---|---|---|---|
| 0.5 | 0.94 | 2062 | 0.35 | 0.77 | 2.19 |
| 0.8 | 0.99 | 2039 | 0.33 | 0.72 | 2.06 |
| 1.0 | 1.00 | 1968 | 0.31 | 0.69 | 1.97 |
| 1.4 | 1.03 | 1987 | 0.31 | 0.67 | 1.93 |
| 1.7 | 1.05 | 1803 | 0.27 | 0.60 | 1.72 |

These conventions span almost an order of magnitude, the well-known ambiguity in mapping a spectral frequency onto a real-space length in a disordered solid; we therefore treat the correlation length as an order-of-magnitude estimate rather than a precise value. What they agree on is that it is of order one nanometer (≈0.3–2 nm); the Ioffe-Regel wavelength (~2 nm) places it within the medium-range regime (0.5–5 nm), between the Si-O coordination shell (~0.16 nm) seen by FTIR/NMR and the ~5 nm colloidal globule of the Jennings CM-II model [9] resolved by small-angle scattering. The boson peak thus accesses the medium-range structural organization of C-S-H, a regime that conventional probes do not reach directly.

This intermediate scale corresponds physically to the spatial extent over which low-frequency collective vibrations maintain coherence before being scattered by structural disorder. In the context of C-S-H, $\xi$ is comparable to the interlayer repeat distance of ideal 14 Å tobermorite (~1.1–1.4 nm), suggesting that the relevant heterogeneity operates at or below the single-layer scale, consistent with the known in-plane disorder of silicate chains and the irregular occupation of $Ca^{2+}$ and $H_2O$ sites within the interlayer.

To place these values in context, Table 7 compares the BP parameters of C-S-H with those of several well-characterized amorphous silicate systems.

Table 7 Boson peak parameters of C-S-H compared with other amorphous silicates

| Material | $\nu_{BP}$ (THz) | $v_t$ (m/s) | $\xi_1$ (nm) | $\xi_2$ (nm) | Method | Ref. |
|---|---|---|---|---|---|---|

| C-S-H (this work) | 0.94–1.05 | 1800–2060 | 0.27–0.35 | 0.60–0.77 | THz-TDS | – |
|---|---|---|---|---|---|---|
| v-$SiO_2$ | ~1.0 | 3760 | ~0.6 | ~1.3 | Raman, INS | [46,55] |
| $SiO_2$-$Al_2O_3$ (7% Al) | ~1.1 | 3860 | ~0.6 | ~1.2 | Raman, THz | [46] |
| $Na_2O$-$SiO_2$ (20% $Na_2O$) | ~1.5 | ~3000 | ~0.3 | ~0.7 | Raman | [56] |
| Soda-lime silicate | ~1.2 | ~3400 | ~0.5 | ~1.0 | Raman | [57] |
| C-S-H (MD, Qomi) | ~1.2–2.5 | – | – | – | MD simulation | [20] |

The BP frequency of C-S-H (~1 THz) falls at the lower end of the range spanned by silicate glasses (1.0–1.5 THz), consistent with its relatively open, water-bearing layered structure and lower shear modulus (~9 GPa vs. ~30 GPa for v-$SiO_2$). The correlation lengths $\xi_1$ and $\xi_2$ are correspondingly shorter than those of dense silicate glasses, reflecting the smaller ratio $v_t/\nu_0$ rather than a fundamentally different structural organization. The key point is that C-S-H behaves, in its low-frequency dynamics, as a member of the same family of disordered silicate networks, not as an anomalous outlier.

A note on the sensitivity of $\xi$ to the assumed shear modulus: the value $G \approx 9$ GPa used here is taken from molecular dynamics simulations [20] and is consistent with nanoindentation measurements on C-S-H gel layers ($G = 8$–$12$ GPa [5]). If G were 50% higher or lower (6–14 GPa), $v_t$ would change by ±22% and $\xi$ would shift proportionally, leaving the order-of-magnitude conclusion unchanged.

**4.4. Debye scaling and the Ioffe–Regel crossover**

A dimensionless measure that allows cross-material comparison is the ratio $\nu_{BP}/\nu_D$, where $\nu_D = k_B\Theta_D/h$ is the Debye cutoff frequency. We estimate the Debye temperature $\Theta_D$ from the mean sound velocity $v_m$ and atomic number density following standard practice [58], obtaining $\Theta_D \approx 288$–$301$ K and $\nu_D \approx 6.0$–$6.3$ THz for the five C-S-H compositions (Table 8).

Table 8 Debye scaling of the boson peak frequency

| Material | $\nu_{BP}$ (THz) | $\Theta_D$ (K) | $\nu_D$ (THz) | $\nu_{BP}/\nu_D$ |
|---|---|---|---|---|
| C-S-H, Ca/Si = 0.5 | 0.94 | 301 | 6.27 | 0.15 |
| C-S-H, Ca/Si = 0.8 | 0.99 | 300 | 6.25 | 0.16 |
| C-S-H, Ca/Si = 1.0 | 1.00 | 296 | 6.17 | 0.16 |
| C-S-H, Ca/Si = 1.4 | 1.03 | 297 | 6.19 | 0.17 |
| C-S-H, Ca/Si = 1.7 | 1.05 | 288 | 6.00 | 0.17 |
| v-$SiO_2$ | ~1.0 | ~495 | ~10.3 | ~0.10 |
| $SiO_2$–$Al_2O_3$ (7% Al) | ~1.1 | ~510 | ~10.6 | ~0.10 |
| $Na_2O$–$SiO_2$ (20% $Na_2O$) | ~1.5 | ~400 | ~8.3 | ~0.18 |
| Soda-lime silicate | ~1.2 | ~420 | ~8.8 | ~0.14 |

The ratio $\nu_{BP}/\nu_D$ for C-S-H (0.15–0.17) sits within the range spanned by silicate glasses (0.10–0.18), toward the high end occupied by modifier-rich compositions such as sodium silicate and soda-lime glass (Fig. 8). This is physically sensible: $Ca^{2+}$ ions in C-S-H play a role analogous to $Na^+$ in sodium silicate glasses: they disrupt the covalent silicate network, soften the structure, and shift the BP to a larger fraction of the Debye frequency. The strong-network end member v-$SiO_2$, with its fully polymerized tetrahedral framework, has the lowest ratio (~0.10). That C-S-H falls closer to the modified glasses than to v-$SiO_2$ is consistent with its partially depolymerized, calcium-rich layered structure.

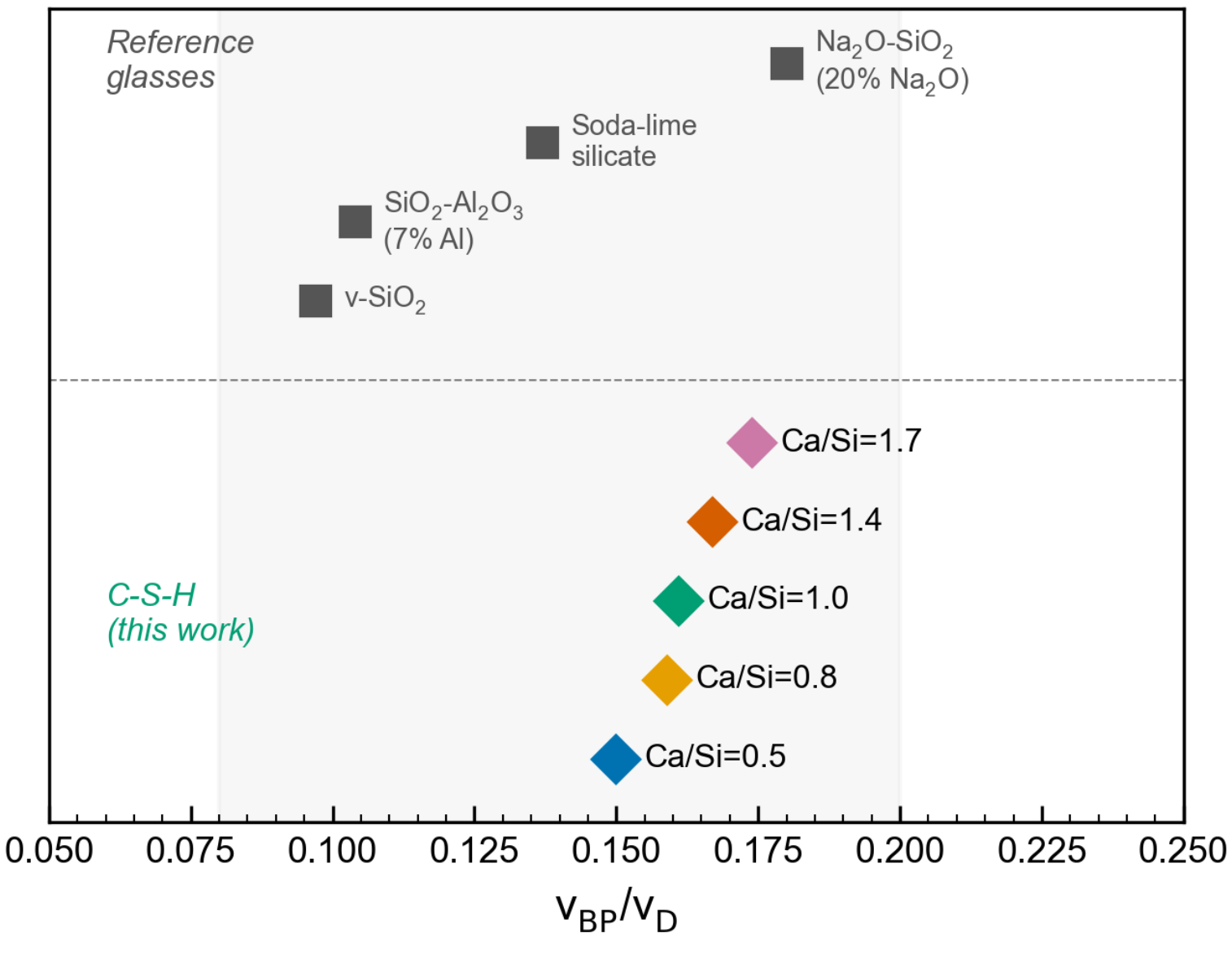

Fig. 8 Debye-normalized boson-peak frequency $\nu_{BP}/\nu_D$ of C-S-H compared with silicate glasses, placing C-S-H within the universal range and toward the modifier-rich end

This scaling analysis carries a broader implication: despite being a hydrated, layered, cement-derived material with no obvious resemblance to a melt-quenched glass, C-S-H obeys the same Debye-normalized BP scaling as conventional silicate glasses. Its low-frequency vibrational anomaly is not a peculiarity of cement chemistry but a manifestation of the universal physics governing disordered silicate networks. The mild increase of $\nu_{BP}/\nu_D$ with Ca/Si (from 0.15 to 0.17) mirrors the trend seen in silicate glasses when the modifier content increases, reinforcing the interpretation that increasing Ca/Si progressively disrupts the silicate network in a manner analogous to alkali modification.

**4.5. Structural interpretation: the Ca/Si ≈ 1.0 crossover**

The convergence of several independent observables at Ca/Si ≈ 1.0 points to a genuine structural crossover in C-S-H:

(i) THz DHO peak height $S \cdot \nu_0/\Gamma$ is maximized and the damping ratio $\Gamma/\nu_0$ reaches a local minimum (this work).

(ii) XRD low-angle diffuse scattering is sharpest (this work).

(iii) TGA structural-water content shows an inflection (this work).

(iv) Bridging-site occupancy in the silicate chains drops to zero (Renaudin et al. [6,59]).

(v) Interlayer calcium occupancy is at its minimum (Renaudin et al. [6]).

(vi) Grangeon et al. [60] reported maximum turbostratic stacking disorder in this Ca/Si range.

Below this threshold, the structural evolution of C-S-H is dominated by progressive silicate chain scission: $Q^2$ sites convert to $Q^1$, the mean chain length shortens, and the silicate backbone loses connectivity. Above it, the dominant process shifts to interlayer space filling by calcium ions and associated water molecules, which gradually stiffen the structure and restore partial medium-range order. The Ca/Si ≈ 1.0 composition marks the point of weakest overall structural constraint, and correspondingly, the richest dipole-active low-frequency dynamics.

From a materials science perspective, this crossover is significant because Ca/Si ≈ 1.0 lies squarely within the range encountered in low-carbon blended cements incorporating supplementary cementitious materials (SCMs) such as fly ash and slag, where the pozzolanic and latent-hydraulic reactions consume calcium and drive the C-S-H to lower

Ca/Si than the ordinary Portland cement value (Ca/Si ≈ 1.7). The boson peak therefore probes the binder phase precisely in the compositional window most relevant to SCM-modified systems; its implications for gauging SCM activity and binder performance are developed in Section 4.7.

### 4.6. Comparison with simulations and prior THz studies

The present results allow a direct assessment of the molecular dynamics prediction by Abdolhosseini Qomi et al. [20], who computed the vibrational density of states of 150 C-S-H models spanning Ca/Si = 1.1–2.1 using the CSH-FF force field. In their reduced VDOS $g(\omega)/\omega^2$, all models exhibited a BP in the low-THz region. Two composition-dependent trends were reported: (i) $\omega_{BP}$ decreased with increasing density, and (ii) the BP intensity $I_{BP}$ increased with increasing density. Since the C-S-H density in their simulations decreased from ~2.55 g/cm$^3$ (Ca/Si ≈ 1.1) to ~2.35 g/cm$^3$ (Ca/Si ≈ 2.1), these trends translate to $\omega_{BP}$ increasing and $I_{BP}$ decreasing with increasing Ca/Si.

Our experimental data confirm both trends qualitatively. The DHO frequency $\nu_0$ increases from 0.94 THz (Ca/Si = 0.5) to 1.05 THz (Ca/Si = 1.7), and the oscillator strength S decreases from 0.038 to 0.016 over the same range. The direction of both trends matches the MD prediction. However, two quantitative discrepancies merit discussion.

First, our experimental $\nu_0$ (0.94–1.05 THz) is systematically lower than the MD range (~1.2–2.5 THz). This likely reflects known limitations of the CSH-FF force field (empirical potentials overestimate low-frequency network stiffness), the finite simulation cells (~2–4 nm, which truncate long-wavelength modes), and the fully-hydroxylated MD models versus our dried samples (interlayer water stiffens the framework and raises $\omega_{BP}$). Despite the offset, both place the BP in the low-THz range, well below Si-O stretching (~15–30 THz), confirming its origin in collective, delocalized motions rather than local bond vibrations.

Second, Qomi et al. correlated $\omega_{BP}$ with density rather than Ca/Si. Our density is non-monotonic (a local maximum at Ca/Si = 1.0; Section3.1) while $\nu_0$ rises monotonically with Ca/Si, indicating that the BP frequency is not set by density alone: the silicate-network topology (chain connectivity, interlayer organization) is an additional variable that density scaling misses, and the tobermorite-templated MD models may not span the topological states of real C-S-H, particularly below Ca/Si ≈ 1.1 where long chains persist.

A necessary control is the role of confined water, the most likely alternative origin of a ~1 THz feature. The CPA decomposition (Section 4.2) isolates the water relaxation as a steep tail (s = 5.4–6.0) that is negligible above ~0.7 THz; quantitatively, the interlayer-water dielectric relaxation lies at 0.1–0.3 GHz, three to four decades below the boson peak, while inelastic neutron scattering places water's vibrational modes (≥1.7 THz) above the 2–30 K $C_p/T^3$ window (Appendix A.2). The ~1 THz loss is therefore the framework boson peak, not confined water, although structural and interlayer water do modulate its damping Γ and strength S and co-vary with Ca/Si. This also accounts for the earlier assignment of Dolado et al. [61], who attributed a ~1 THz feature in hydrated cement pastes to confined water: the difference is one of sample state, their pastes retaining abundant pore water that overlaps the framework feature, whereas our pure, dried, impurity-corrected C-S-H isolates it.

These values pertain to a controlled, partially-dried state (~15–18 wt% retained water, Table 1; well below the ~27 wt% of fully-saturated C-S-H), chosen deliberately to fix the water content across the series so that the compositional trends and Debye-normalized scaling are not confounded by variable hydration. In fully-saturated C-S-H the absolute $\nu_0$, Γ, and S would shift, but the framework boson peak and its compositional behaviour are robust to hydration state.

The non-monotonic DHO damping ratio $\Gamma/\nu_0$, with a local minimum at Ca/Si ≈ 1.0, together with the maximum in DHO peak height at that composition, constitutes a new experimental benchmark that has no counterpart in the existing MD literature. Reproducing this non-monotonic behavior, which reflects the competition between chain depolymerization and interlayer stiffening, would be a stringent test of whether a given force field or ab initio method captures the medium-range structural heterogeneity of C-S-H correctly.

### 4.7. What the boson peak reveals about C-S-H

*A new, dynamical structural dimension.* Beyond confirming the BP, these findings add what conventional characterization could not. The BP-derived correlation length ξ of order 1 nm (≈0.3–2 nm) identifies a medium-range dynamical coherence scale, the distance over which collective vibrations stay in phase before being scattered, that sits between the ~0.16 nm Si-O coordination shell (NMR, FTIR) and the ~5 nm colloidal globule (SAXS/SANS) and had not previously been resolved by experiment in C-S-H; its existence shows that C-S-H is not simply "amorphous" beyond the first coordination shell. Disorder in C-S-H is moreover not a single axis: $\xi$ varies by only ~20% across Ca/Si = 0.5–1.7 while the damping $\Gamma$ and strength S change two-fold, so the dynamically correlated domains stay roughly constant in size while the severity of disorder within them changes dramatically, an independent dynamical dimension that static descriptors (mean chain length, layer spacing) do not capture. This nanometer-scale elastic heterogeneity is also finer than, and complementary to, the mesoscale modulus dispersion familiar in cement science: colloidal and micromechanical models resolve C-S-H into lower- and higher-density populations of differing stiffness at the ~100 nm packing scale [9], whereas $\gamma$ quantifies the stiffness fluctuation within the C-S-H solid itself, below the resolution of any mechanical test, and is a descriptor that only the boson peak makes accessible. Finally, the maximum in BP peak height and minimum in damping at Ca/Si ≈ 1.0 give the first *dynamical* signature of calcium's dual structural role (removing bridging tetrahedra at low Ca/Si, filling and stiffening the interlayer at high Ca/Si [6,59,60]), a crossover previously documented only by static probes ($^{29}$Si NMR chain length, XRD basal spacing, TGA water binding).

*Connection to thermal transport.* The boson peak is the spectroscopic fingerprint of the same low-frequency, diffusive vibrations that govern heat conduction in disordered solids [40], so the parameters obtained here can be carried directly into a thermal-conductivity estimate. Treating C-S-H in the Cahill-Pohl minimum-conductivity limit [62] with the transverse sound velocity inferred from the boson peak ($v_t \approx 1.8$–$3.2$ km s$^{-1}$, depending on whether the gel-layer or intrinsic shear modulus is used) and the measured atomic number density yields $\kappa \approx 0.8$–$1.0$ W m$^{-1}$ K$^{-1}$ at 300 K, essentially independent of Ca/Si. This estimate agrees with equilibrium molecular-dynamics values for C-S-H (~1.0 W m$^{-1}$ K$^{-1}$) [20] and with transient-plane-source measurements on hardened cement pastes once porosity is accounted for [41].

The near composition-independence has a clear physical origin. Zhou et al. [63] decomposed heat transport in C-S-H into propagating (propagon, ≳30%) and non-propagating diffusive (diffuson, ≈70%) contributions, the latter described by Allen-Feldman theory. The diffuson contribution is pinned at the glassy minimum and is, by construction, insensitive to the degree of disorder, which is why both the present Cahill-Pohl estimate and the molecular-dynamics value vary little with Ca/Si. The propagon contribution, in contrast, is limited by Rayleigh scattering from the medium-range elastic heterogeneity, precisely the variance $\gamma$ that the boson peak makes accessible here, since the propagon scattering rate scales with $\gamma$. Because $\gamma$ falls roughly two-fold from Ca/Si = 0.5 to 1.7 (Table 2), the boson peak supplies the elastic-heterogeneity input that should govern the propagon-borne part of $\kappa$; how strongly this modulates the minor propagon contribution, and hence any residual composition dependence of the thermal conductivity, is a target for dedicated thermal-transport studies. Thermal transport in C-S-H is thus governed not by porosity and moisture alone, as commonly assumed, but also by the vibrational character of the solid framework.

*Connection to creep.* The most consequential implication concerns creep, the time-dependent deformation under sustained load that is concrete's single most important long-term mechanical response. Abdolhosseini Qomi et al. [5] showed by MD that the logarithmic creep of C-S-H arises from stress-driven rearrangements of the disordered calcium-silicate network, governed by its low-frequency vibrational landscape, and in glass physics the BP and the α-relaxation underlying creep are increasingly regarded as the harmonic and anharmonic faces of the same population of quasi-localized modes [64,65]. The CPA disorder parameter $\gamma$ quantifies the spatial variance of the elastic modulus, precisely the heterogeneity that sets the distribution of energy barriers for local rearrangements, so a higher $\gamma$ implies a broader barrier distribution and potentially higher creep compliance. This offers a route to a standing problem: Hu et al. [66] found that higher-Ca/Si C-S-H creeps less but attributed this mainly to non-creeping portlandite rather than to the C-S-H itself. Because our $\gamma$ is obtained after EMT removal of crystalline impurities (including portlandite), it reflects the intrinsic C-S-H alone, and its monotonic decrease from 0.98 (Ca/Si = 0.5) to 0.48 (Ca/Si = 1.7) predicts that the intrinsic creep susceptibility should fall with Ca/Si independently of portlandite content, a falsifiable hypothesis testable by single-phase nanoindentation creep on the same samples.

*Implications for low-carbon binders.* The boson peak is a generic signature of medium-range disorder in glasses and other amorphous solids, and the SCMs central to low-carbon cements (fly ash, slag, calcined clays, natural pozzolans) are themselves largely amorphous. The approach demonstrated here for C-S-H therefore extends directly to the SCMs themselves: their boson peak provides a structure-based, non-destructive descriptor of the amorphous network, which we propose, as a falsifiable hypothesis, correlates with their pozzolanic reactivity, complementing the empirical indices (amorphous content, fineness, reactivity tests) conventionally used to grade SCMs. Quantifying this correlation is the subject of ongoing work.

Together, these connections show that the BP is not a spectroscopic curiosity but a gateway to the low-frequency dynamics that govern the rate-dependent thermal and viscoelastic performance of cementitious binders and the compositional design of their low-carbon variants.

## 5. Conclusions

(1) THz-TDS measurements on synthetic C-S-H with Ca/Si = 0.5–1.7, after correction for crystalline impurities using Bruggeman effective medium theory, reveal a broad dielectric absorption feature near 1 THz in the frequency-normalized loss spectra of all samples. DHO fitting confirms this feature as the boson peak, with a characteristic frequency consistent with prior molecular dynamics predictions.

(2) Low-temperature heat capacity measurements independently confirm the BP through the observation of a $C_p/T^3$ excess peak at 7–12 K. After correcting for the known systematic offset between calorimetric and spectroscopic BP frequencies, the two probes agree on a characteristic energy scale of ~1 THz.

(3) DHO decomposition reveals that the integrated spectral weight S decreases monotonically with Ca/Si, tracking the calorimetric $C_p/T^3$ trend, while the peak height $S \cdot \nu_0/\Gamma$ reaches a maximum at Ca/Si ≈ 1.0, where low damping combines with still-substantial oscillator strength. The apparent "strongest BP" at Ca/Si ≈ 1.0 in the raw spectra reflects not the largest number of excess modes, but the most coherent, least damped, and most dipole-active ones. This composition coincides with the structural crossover from silicate chain depolymerization to interlayer calcium filling.

(4) The boson peak yields a medium-range dynamical correlation length of order 1 nm ( ≈ 0.3–2 nm), at or below the single-layer scale of C-S-H and distinct from the ~5 nm colloidal packing unit of the CM-II model. The weak dependence of $\nu_0$ on Ca/Si indicates that composition modulates the severity of disorder within domains of similar spatial extent, rather than the domain size itself.

These results establish the boson peak as an experimentally accessible probe of medium-range structure in C-S-H, operating at a length scale between the short-range coordination environment visible to NMR/FTIR and the mesoscale organization captured by small-angle scattering. Conceptually, the boson peak supplies a medium-range *dynamical* coherence scale that static probes cannot reach, and an elastic-heterogeneity parameter $\gamma$ that further predicts the composition dependence of the intrinsic thermal transport and creep of the binder. The combination of THz spectroscopy and low-temperature calorimetry provides complementary views, one sensitive to dipolar dynamics, the other to the full vibrational spectrum, that together yield a richer picture of the low-frequency excitations in this disordered material than either technique alone. More generally, combining a direct spectroscopic probe with calorimetry is an established route to the boson peak across glass families, from oxide [43] to molecular glasses [67]; our results extend this two-probe strategy to the regime of broad, heavily damped boson peaks characteristic of hydrated and strongly disordered amorphous solids, where the calorimetric inverse problem alone cannot resolve the peak frequency and the spectroscopic probe becomes essential.

## Acknowledgments

This work was supported by the National Natural Science Foundation of China (Grant No. 52178238).

## Declaration of competing interest

The authors declare that they have no known competing financial interests or personal relationships that could have appeared to influence the work reported in this paper.

## CRediT authorship contribution statement

**Xiangyu Li**: Conceptualization, Formal analysis, Writing – original draft, Funding acquisition. **Ying Chen**: Investigation (sample synthesis and THz-TDS measurements). **Jipeng Luo**: Investigation (low-temperature heat-capacity/PPMS measurements). **Ya Chen**: Investigation (THz-TDS), Formal analysis. **Linhao Wang**: Investigation (XRD and FTIR characterization). **Lidan Tian**: Investigation (sample preparation), Formal analysis. **Gan Ding**: Formal analysis (spectral fitting). **Zeyu Lu**: Formal analysis, Writing – review & editing. **Zhangli Hu**: Investigation (sample synthesis), Writing – review & editing. **Biqin Dong**: Writing – review & editing. **Yue Li**: Writing – review & editing. **Zongjin Li**: Supervision, Writing – review & editing.

## Data availability

The data that support the findings of this study are available from the corresponding author upon reasonable request.

## Appendix A. Supplementary analyses

### Appendix A.1. Comparison of calorimetric boson-peak fitting models

To examine whether the BP frequency can be extracted directly from the $C_p$ data, we fitted the low-temperature heat capacity to five models: (i) Debye + Gaussian-distributed Einstein oscillators, (ii) Debye + log-normal Einstein oscillators, (iii) Tikhonov-regularised numerical inversion of the vibrational density of states [46], (iv) a forward model driven by the THz-derived DHO spectral function, and (v) a Planck-like parametrisation of the SPM soft-mode VDOS adapted from Ding et al. [45]. All five reproduce the measured heat capacity comparably well but differ markedly in the BP frequency they return. We assess fit quality from the agreement in the figures rather than from tabulated $R^2$: because $C_p$ is dominated by the high-temperature Debye contribution, a high $R^2$ on $C_p$ is routine and says little about the boson-peak region.

Methods (i) and (ii) give Einstein temperatures differing by up to a factor of two for the same sample (e.g., Ca/Si = 1.4: $\theta_E$ = 19.1 K vs 35.0 K) despite indistinguishable fit quality; converted to frequency these correspond to $\nu_E$ = 0.40–0.73 THz, well below the THz DHO values (0.94–1.05 THz). Method (iii) reproduces $C_p(T)$ yet yields a monotonically decreasing $g(\nu)/\nu^2$ with no resolvable peak. These three methods are compared in Fig. A.5. The ambiguity arises because the $C_p(T)$ integral is dominated by the total excess mode count rather than the frequency distribution; different VDOS shapes with the same integrated weight produce nearly identical $C_p(T)$ curves.

Method (v), the physically motivated choice retained in the main text, decomposes the heat capacity as $C_p = \gamma T + n_D C_{Debye}(\Theta_D) + n_q C_{quasi}$, with the quasi-localised contribution computed from the Planck-like reduced VDOS $g_q(\nu)/\nu^2 = \varphi \nu^3/(e^{\ell\nu}-1)$, whose low-frequency limit $g_q(\nu) \sim \nu^4$ matches the universal scaling of quasi-localised modes [45]. A four-parameter fit ($\gamma$, $\Theta_D$, $n_q/3N$, $\ell$) over 2–30 K reproduces the data most closely of all five methods. As the upper fit cutoff moves from 30 to 50 K, the nuisance $\Theta_D$ rises from ~620–670 K to ~720–750 K (the higher-temperature data sample modes outside the Debye-plus-soft-mode description) while $\nu_{BP}$ changes by less than 5%; we adopt 30 K to isolate the BP region.

For Method (iv) we further tested whether a single SPM soft-mode VDOS could simultaneously reproduce the THz $\varepsilon''(\nu)/\nu$ lineshape and the $C_p/T^3$ curve under a frequency-dependent coupling $C(\nu) \propto \nu^q$: the two peak positions are reconciled by a consistent exponent ($q \approx -0.4$ across all compositions), but the full lineshapes cannot be unified, because the THz dielectric response (a damped-oscillator profile) and the bare vibrational density of states (an asymmetric $\omega^4$ profile) have intrinsically different shapes. The two probes therefore agree on the boson-peak energy while differing in lineshape, consistent with the coupling-weighted intensity decoupling reported in Section 3.

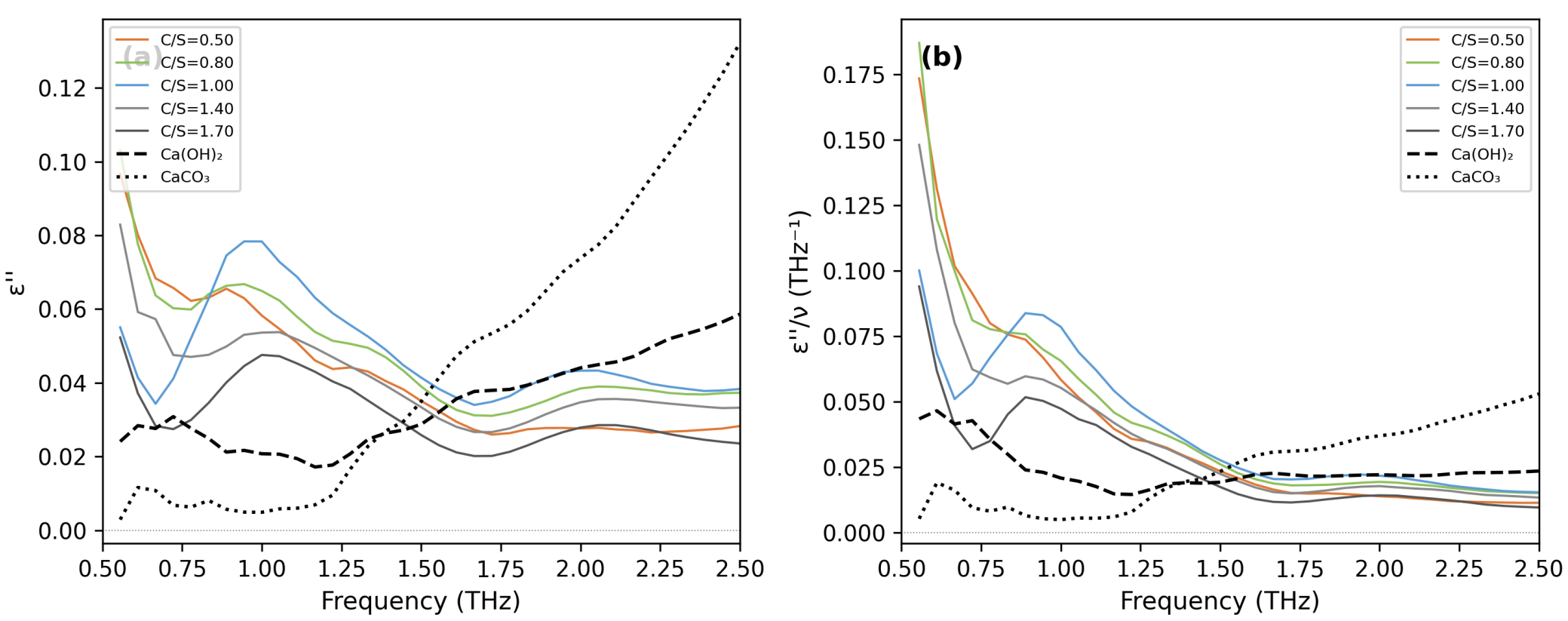

Fig. A.1 Dielectric loss of crystalline $Ca(OH)_2$ and $CaCO_3$ compared with C-S-H. Neither crystalline phase exhibits a boson-peak feature in $\varepsilon''(\nu)/\nu$

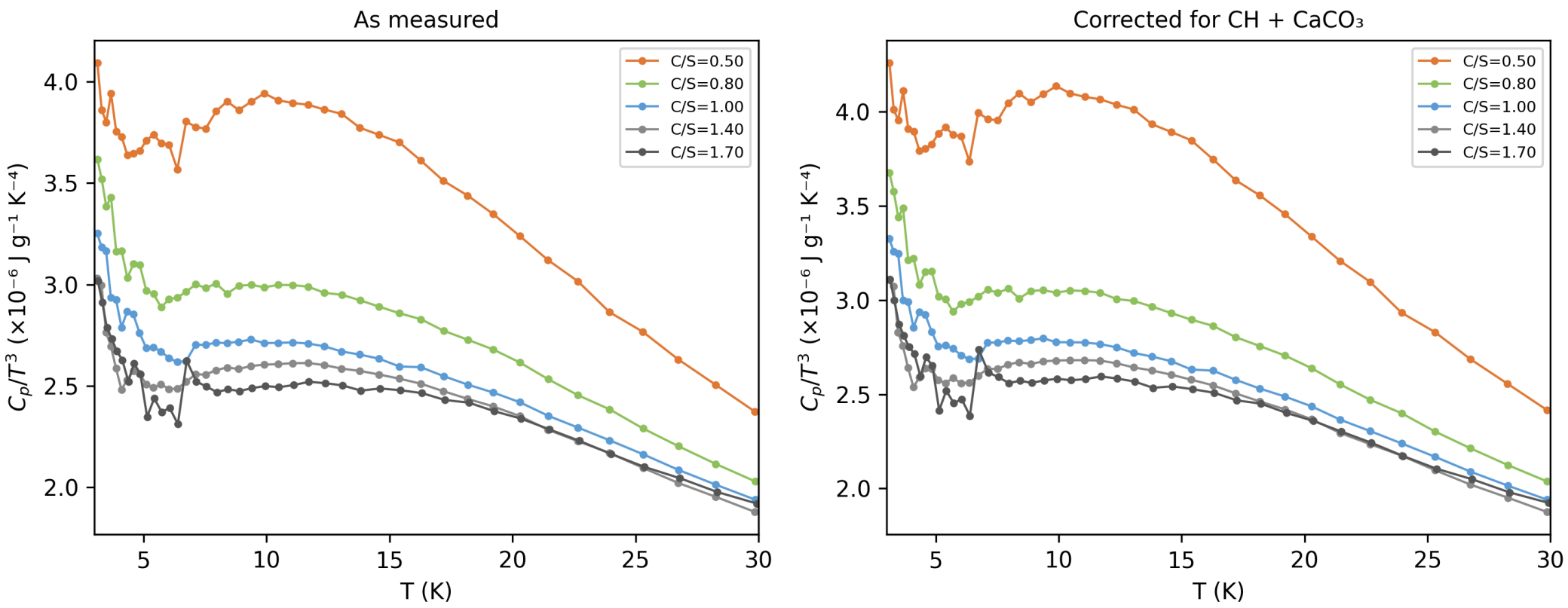

Fig. A.2 Effect of crystalline-impurity correction on the $C_p/T^3$ data: as-measured versus corrected for $Ca(OH)_2$ and $CaCO_3$. The correction alters peak heights by <5% and preserves the compositional trend

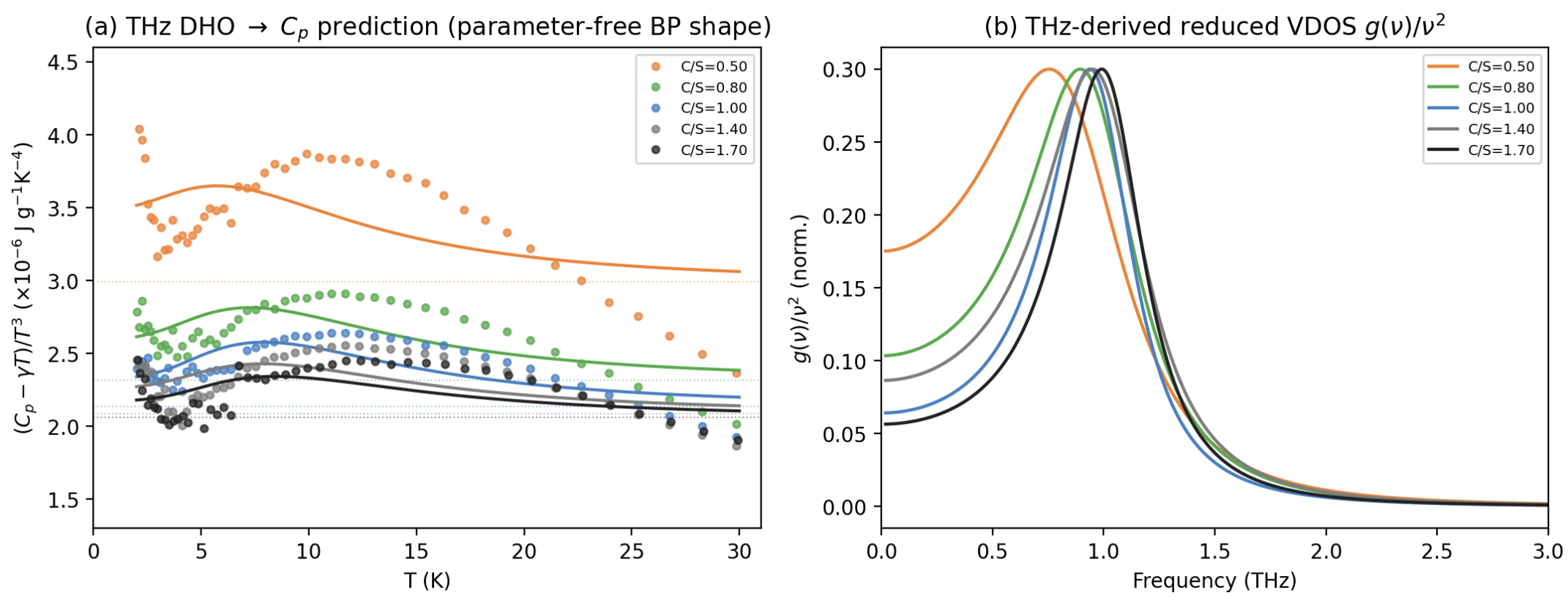

Fig. A.3 Forward prediction of $C_p/T^3$ from the THz-derived DHO spectral function, with the boson-peak frequency and width fixed by THz (no free frequency parameters). The parameter-free prediction reproduces the calorimetric excess for four of the five compositions

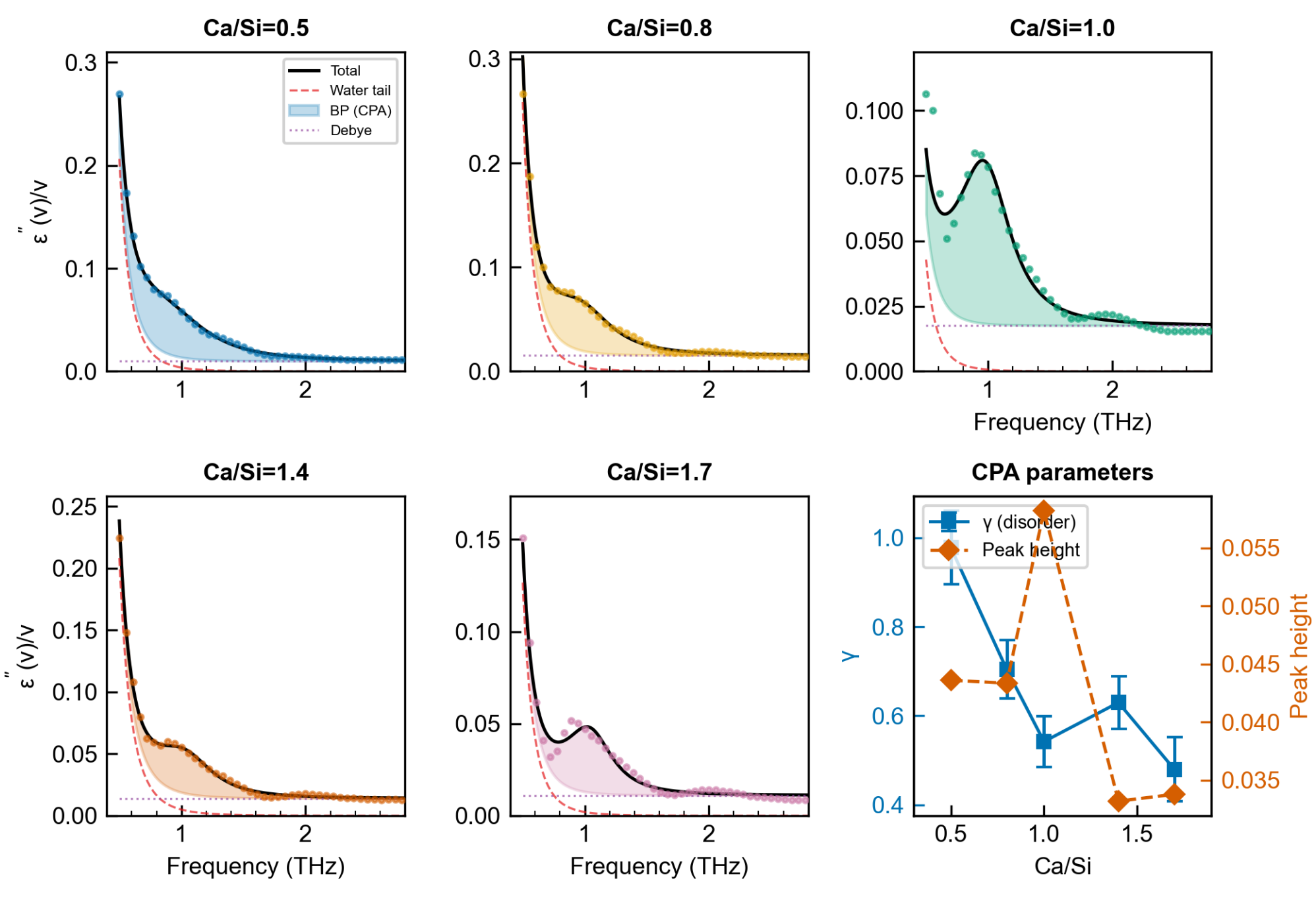

Fig. A.4 CPA multi-physics superposition fits of ε″(ν)/ν, decomposing each spectrum into the water-relaxation tail, the CPA-DHO boson peak (governed by the disorder parameter γ), and the Debye baseline

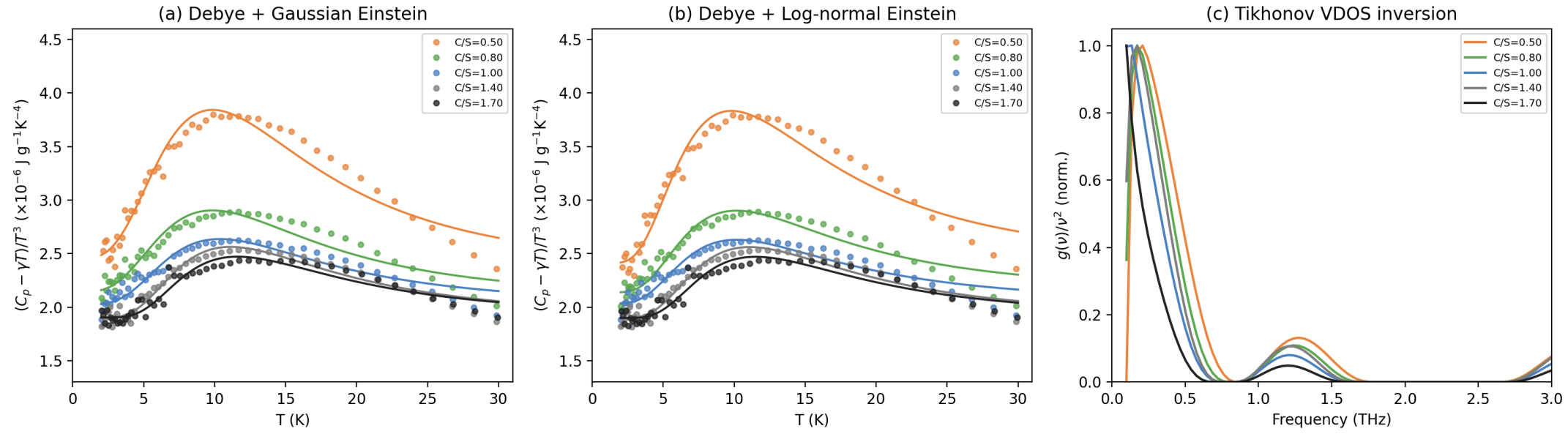

Fig. A.5 Comparison of three calorimetric fitting methods: (a) Debye + Gaussian Einstein, (b) Debye + log-normal Einstein, (c) Tikhonov-regularized VDOS inversion. All reproduce $C_p$ yet return non-unique or unresolved boson-peak frequencies

**Appendix A.2. Quantitative bounds on the interlayer-water contribution**

The interlayer-water dielectric relaxation frequencies quoted in Section 4.6 follow from the relaxation-time–basal-spacing relation of ref. [52], $\tau_{cc}(d) = A \exp[-(d-A_0)/h] + \tau_\infty$ (with $A = 12655$ ps, $A_0 = 3.33$ Å, $h = 2.90$ Å, $\tau_\infty = 14$ ps), evaluated at the basal spacing $d_{001}$ extracted from the low-angle X-ray reflection of each sample; the resulting values are collected in Table A.1. Even adopting the maximum relaxation time reported in ref. [52] (~0.5 ns) gives a conservative ≥0.3 GHz, still 3.5 decades below the boson peak.

Table A.1 Basal spacing and interlayer-water dielectric relaxation of C-S-H, computed from the Cole–Cole model of ref. [52]

| Ca/Si | $D_{001}$ (nm) | $\tau_{cc}$ (ns) | $f_{relax}$ (GHz) | Decades below 1 THz |
|---|---|---|---|---|
| 0.5 | 1.01 | 1.23 | 0.13 | 3.9 |
| 0.8 | 0.96 | 1.46 | 0.11 | 4.0 |
| 1.0 | 1.17 | 0.73 | 0.22 | 3.7 |
| 1.4 | 1.01 | 1.24 | 0.13 | 3.9 |
| 1.7 | 0.98 | 1.36 | 0.12 | 3.9 |

On the calorimetric side, the bound on water modes follows from inelastic neutron scattering [68], which places water's translational modes at 1.7–9 THz and librations at 9–30 THz, above the boson peak; since an Einstein mode of frequency ν peaks in $C_p/T^3$ near $T \approx 9.7\nu$ (THz) K, the lowest water band contributes a maximum near 16 K, above the ~9 K $C_p/T^3$ peak, leaving the 2–30 K window dominated by the solid framework.